%% file: main.tex
\documentclass[superscriptaddress,pra,aps,twocolumn,10pt,longbibliography]{revtex4-2}

\usepackage{graphicx}
\usepackage{dcolumn}
\usepackage{bm}
\usepackage{physics} 
\usepackage{amsfonts} 
\usepackage{amsthm}
\usepackage{graphicx} 
\usepackage{bm} 
\usepackage[colorlinks=true,linkcolor=black, urlcolor=blue, citecolor=purple]{hyperref} 
\usepackage{tabularx}
\usepackage[table]{xcolor}
\definecolor{tableShade}{gray}{0.9}
\newcolumntype{a}{>{\hsize=.75\hsize}X}
\newcolumntype{b}{>{\hsize=2\hsize}X}

\newcommand{\specialcell}[2][c]{%
  \begin{tabular}[#1]{@{}c@{}}#2\end{tabular}}

\usepackage{glossaries}
\loadglsentries{glossary.tex}

\DeclareMathOperator{\EX}{\mathbb{E}}

\newcommand{\mb}[1]{\mathbf{#1}}

\newcommand{\schro}{Schr{\"o}dinger }

\usepackage{empheq}
\usepackage{xcolor}
\usepackage[most]{tcolorbox}
\colorlet{shadecolor}{orange!15}
\theoremstyle{definition}

\begin{document}


\preprint{APS/123-QED}

\title{Wave function Ansatz (but Periodic) Networks  \\and the Homogeneous Electron Gas} 

\author{Max Wilson}
\email{amawi@dtu.dk}
\affiliation{Technical University of Denmark, Department of Applied Mathematics and Computer Science, Kgs. Lyngby, Denmark}
\author{Saverio Moroni}%
\affiliation{Consiglio Nazionale delle Ricerche, Istituto Officina dei Materiali and Scuola Internazionale Superiore di Studi Avanzati, Trieste, Italy}
\author{Markus Holzmann}
\affiliation{Univ. Grenoble Alpes, CNRS, LPMMC, 38000 Grenoble, France} 
\author{Nicholas~Gao}
\affiliation{Technical University of Munich, Munich, Germany}
\author{Filip Wudarski}
\affiliation{Quantum Artificial Intelligence Lab. (QuAIL), Exploration
Technology Directorate, National Aeronautics and Space Administration Ames Research Center,
Moffett Field, CA 94035, USA}
\affiliation{Universities Space Research Association, Research Institute for Advanced Computer Science, Mountain View, CA 94043, USA}
\author{Tejs Vegge}
\affiliation{Technical University of Denmark, Department of Energy Conversion and Storage, Kgs. Lyngby, Denmark}
\author{Arghya Bhowmik}
\affiliation{Technical University of Denmark, Department of Energy Conversion and Storage, Kgs. Lyngby, Denmark}

\date{\today}

\begin{abstract}
\noindent 
We design a neural network Ansatz for variationally finding the ground-state wave function of the Homogeneous  Electron Gas, a fundamental model in the physics of extended systems of interacting fermions.  We study the spin-polarised and paramagnetic phases with 7, 14 and 19 electrons over a broad range of densities  from $r_s=1$ to $r_s=100$, obtaining similar or higher accuracy compared to a state-of-the-art  iterative backflow baseline even in the challenging regime of very strong correlation. Our work extends previous applications of neural network Ans\"{a}tze  to molecular systems with methods for handling periodic boundary  conditions, and makes two notable  changes to improve performance: splitting the pairwise streams  by spin alignment and generating backflow coordinates for the orbitals from the network. We illustrate the advantage of our high quality wave functions in computing the reduced single particle density matrix. This contribution establishes neural  network models as flexible and high precision Ans\"{a}tze for periodic electronic systems, an important step towards applications to crystalline solids.
\end{abstract}

\maketitle

\section{Introduction}\label{sec:introduction}

Electronic structure theory forms the backbone of ab-initio calculations of systems in quantum chemistry and condensed matter physics. Method development of approximate solutions to the many-electron \schro equation is a central request of theory, pushed further with algorithmic improvements and the continuing growth of computational resources \cite{martin2020electronic}. Approximate solutions to the electronic \schro equation together with advanced high-throughput, automated methods have broad applications ranging from pharmaceutical design to new materials for clean energy technologies like batteries, and machine learning methods are promising to address the various challenges involved.

\begin{figure*}
    \centering
    \includegraphics[width=0.9\textwidth]{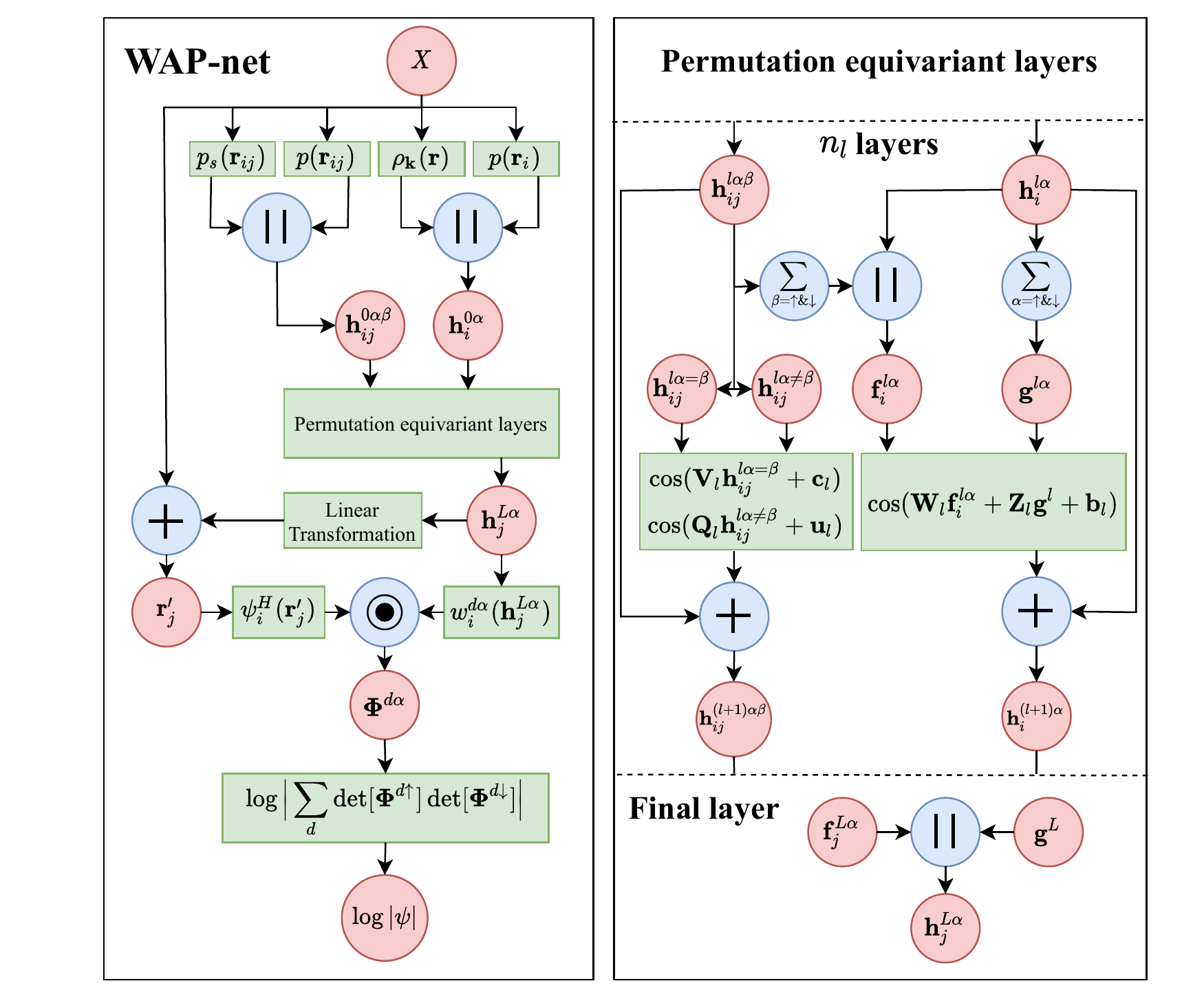}
    \caption{Diagram showing key functions (green), variables (red), and operations (blue) in WAP-net. All functions and operations are described in Section~\ref{sec:methods} (Ansatz). Roughly, system configurations ($X$) are embedded by periodic functions ($p_s(\cdot)$, $p(\cdot)$,  $\rho_\mb{k}(\cdot)$), split into single and pairwise streams, and passed through equivariant layers. The resulting variables are used to generate orbitals in Slater determinants that compose the log-amplitude of the Ansatz. In the operations, $||$ represents a concatenation, $+$ the addition, and $\bullet$ the product. In an abuse of notation the \& symbol in the sums indicates the sums are taken over both $\uparrow$ and $\downarrow$ separately. The diagram is representative but not exact, it does not contain all operations, for an exact description see Section~\ref{sec:methods}.
    \label{fig:wapnet}}
\end{figure*}

In this paper, we develop a periodic neural network Ansatz, dubbed \gls{wap}-net and outlined in Figure~\ref{fig:wapnet}, to describe the ground-state of the \gls{heg} over a broad density regime. The model is composed of permutation equivariant layers \cite{pfau2019ab}, products of Hartree-Fock plane-wave orbitals appropriate to a homogeneous system and prefactors \cite{hermann2019deep}, periodic input features \cite{pescia2021neural}, and backflow coordinates \cite{OBF}. Additionally, improvements are made to previous related networks by splitting pairwise interaction data by spin alignment and including orbital backflow coordinates. We compare the obtained \gls{wap}-net energies to those from \gls{ib} network wave functions \cite{taddei2015iterative} and show that our method provides the most accurate explicit wave functions in almost all cases and often competes with the energies obtained from fixed-node \gls{dmc} calculations \cite{martin2016interacting} performed on the top of our best \gls{ib} wave function for the \gls{heg}.

The \gls{heg}, also known as Jellium or the Uniform Electron Gas \cite{mahan2000homogeneous}, is a simplified model allowing to capture intricate properties of solid systems such as the quantum nature of delocalized electrons without requiring the introduction of a specific atomic lattice \cite{jacobsen1987interatomic, foulkes2001quantum, harrison1980solid, giuliani2005quantum,martin2016interacting}. It has played a prominent role in the development of theoretical approaches such as diagrammatic perturbation theory \cite{Hedin65,martin2016interacting,chen2019combined}, \gls{qmc} \cite{ceperley1978ground,ceperley1980ground}, and, more recently, in the exportation of quantum chemistry methods to extended systems \cite{shepherd2012investigation, shepherd2013many,ruggeri2018correlation}. In all of these approaches, the \gls{heg} has been used as a first milestone on the way towards more accurate description of electronic correlations in materials and is a natural first step to developing novel neural network Ans\"{a}tze for solids.

Methods involving neural networks have already demonstrated state-of-the-art performance in problems spanning fields including physics \cite{raissi2019physics}, chemistry \cite{pfau2019ab}, games \cite{silver2017mastering}, and autonomous systems \cite{kiumarsi2017optimal}, driven by rapid evaluation on GPUs, availability of data, and design innovations. First established as an Ansatz for quantum many-body systems in 2017 \cite{carleoSolvingQuantumManyBody2017}, subsequent methods introduced other ideas from the rapidly developing field of neural networks. Recent attempts have included physics based structure into the Ans\"{a}tze, where more sophisticated techniques in network design and optimisation have even advanced this subfield to the state-of-the-art performance on molecular systems \cite{pescia2021neural, hermann2019deep, pfau2019ab, wilsonSimulationsStateoftheartFermionic2021, gao_pesnet_2022, spencer2020better}.

Recent diagrammatic Monte Carlo calculations \cite{Dia} have benchmarked values of the renormalization factor, $Z$, characterizing the Fermi liquid behavior of the HEG, against previous quantum Monte Carlo results
\cite{momk3D}. Here we use WAP-net to calculate the reduced single particle density matrix and
estimate the possible fixed-node bias in Ref.~\cite{momk3D}.

The remainder of this article is organised as follows: Section~\ref{sec:related_work} highlights relevant literature and its relationship to this research, Section~\ref{sec:methods} gives a complete theoretical description of the method, including hyperparameters and setup, Section~\ref{sec:results} outlines the results,
first on the ground state energy (\ref{subsec:energy})
needed to judge the performance of WAP-net
(\ref{subsec:discussion}).  Then  we apply our wave functions to  calculate the reduced single particle density matrix (\ref{subsec:obs}). Finally, the paper is summarised and next steps are proposed in Section~\ref{sec:conclusion}.

Since a preprint of this research was first made available multiple groups have made grand strides in the application of neural networks to quantum Monte Carlo, including self-attention networks \cite{von2022self} and extended systems \cite{li2022ab}, further establishing the importance, flexibility, and potential of these methods. 

\section{Related work}\label{sec:related_work}

Regarding fermionic systems, the accuracy of variational \gls{qmc} methods are, in general, ultimately limited by the underlying form of the trial wave function (the Ansatz). In standard implementations, antisymmetry is guaranteed by one or several Slater determinanats constructed from single particle orbitals that are correlated via an explicit pairwise bosonic (permutation invariant) potential, frequently called the Slater-Jastrow wave function. Backflow wave functions further correlate the particle positions entering the orbitals of the Slater determinants \cite{Schmidt81}, or more general permutation equivarant forms \cite{BFN,caffarel10,OBF}, from which \gls{ib} networks \cite{taddei2015iterative, ruggeri2018nonlinear}  currently provide the most accurate systematic improvement in homogeneous systems \cite{BFStoner}.

Research into neural network Ans\"{a}tze has covered Boltzmann machines and feedforward neural networks over a wide range of systems \cite{carleoSolvingQuantumManyBody2017, kessler2021artificial,hanSolvingManyelectronSchrodinger2019,chooFermionicNeuralnetworkStates2020}, including more recent Slater determinant Ans\"{a}tze \cite{acevedoVandermondeWaveFunction2020, liNeuralnetworkbasedMultistateSolver2021}. Other more recent examples incorporate physics based structure \cite{hermann2019deep}, or large and deep networks \cite{pfau2019ab, spencer2020better, wilsonSimulationsStateoftheartFermionic2021} on molecular systems. Associated implementations such as the PESNet attempt to curtail the learning process of a potential energy surface \cite{gao_pesnet_2022} and the neural canonical transformation for the computation of the electron effective mass \cite{wang2022m}. Finally, there is work on incorporating periodic functions to neural networks and/or applying these methods to periodic data/systems \cite{ziyin2020neural, sitzmann2020implicit,Pescia} and equivariances to periodic data \cite{schutt2018schnet, xie2021crystal}.

In this contribution, we focus on a \gls{qmc} neural network Ansatz method applied to the \gls{heg} model. Other recent work tackles the  problem of designing a neural network Ansatz for bosonic systems on a torus \cite{pescia2021neural}, resulting in some similarities in function design such as the application of periodic embedding functions to the electron coordinates. Quantum chemistry benchmarks on the \gls{heg} \cite{shepherd2012investigation,shepherd2013many,ruggeri2018correlation} are limited to the high density region and small number of electrons. Since neural networks are highly parameterisable  and can model arbitrary smooth functions \cite{hornik1989multilayer}, one can expect that this computational model in conjugation with \gls{qmc} techniques will be flexible enough to approach high accuracy uniformly in density, similar or better than \gls{ib} Ansatz. Additionally, it may potentially provide a more favourable scaling (for a given accuracy) with electron number than other quantum chemistry methods. 

\section{Methods}\label{sec:methods}

\subsection*{Hamiltonian}

In the following we consider a system of $N_e$ electrons in three spatial dimensions and seek the ground-state wave function $\psi(X)$ of the time-independent \schro equation
\begin{equation}
    H \psi(X) = E \psi(X),
    \label{eq:schrodinger}
\end{equation}
where $X = (\mb{r}_1,\dots,\mb{r}_{N_e})$ denotes a full configuration of the electron coordinates $\mb{r}_i=(r_{i,x},r_{i,y},r_{i,z})$.
In atomic units, the Hamiltonian 
\begin{equation}
    H = -\frac{1}{2}\nabla_X^2 + V(X)
    \label{eq:hamiltonian}
\end{equation}
contains the kinetic energy operator $\nabla_X^2$ of each electron where $\nabla_X = (\nabla_{\mb{r}_1},\dots, \nabla_{\mb{r}_{N_e}})$ and
\begin{align}
    \nabla_X^2 = \sum^{N_e}_{i=1} \Big(\frac{\partial^2}{\partial r_{i,x}^2} + \frac{\partial^2}{\partial r_{i,y}^2} + \frac{\partial^2}{\partial r_{i,z}^2} \Big).
    \label{eq:kinetic}
\end{align}
The potential energy is $V(X)$, a function of pairwise interactions $v(\mb{r})$ depending on the distance between two particles
$\mb{r}_{ij} = \mb{r}_j-\mb{r}_i$,
\begin{equation}
V(X)=\sum_{i<j}v(\mb{r}_{ij}).\
\end{equation}
In order to describe the bulk of a extended systems of electron density $\rho_e=N_e/L^3$, with electrons in a cubic box of side $L$, we want to impose periodic boundary conditions, e.g. $\psi(\dots,r_{i,x} +L,\dots) = \psi(\dots, r_{i, x,\dots})$.
Since the Coulomb interaction is long ranged, some care is needed for proper setup and computation of the interaction potential. Here, we use the standard Ewald summation \cite{Ewald,martin2020electronic} in direct and reciprocal space for its computation given charges $q_i$ and $q_j$, for electrons in atomic units $q_i=1$,
\begin{align}
    v(\mb{r}) &= q_iq_j \Bigg[ \sum_{\mb{T}} \frac{ \text{erfc}[\kappa ||\mb{r} - \mb{T}||]}{||\mb{r} - \mb{T}||} 
- \frac{1}{\kappa^2 L^3} \nonumber
    \\ &+  \frac{4\pi}{L^3}\sum'_{\mb{k}} \frac{ \exp(-\mb{k}^2/(4\kappa^2))}{\mb{k}^2} \exp(-i\mb{k} \cdot \mb{r}) \Bigg]
\end{align}
where the first summation is over all the lattice vectors to image cells, $\mb{T}=(i_x,i_y,i_z) L$ with integers $i_\alpha$, whereas the second summation is over all reciprocal lattice vectors $\mb{k}=2 \pi (i_x,i_y,i_z)/L$ and the prime indicates the omission of $\mb{k}=(0,0,0)$ term. The convergence of both summations is determined by the hyperparameter $\kappa$ which should be set to minimize the size of the sets required to make the sums converge. 

For convenience, we also add the interaction energy of each electron with its own images, in our case for a system of $N_e$ electrons it is $N_e v_M$, to the total interaction energy $V(X)$ where 
\begin{align}
v_M = \lim_{\mb{r}_{ij} \to 0} \left[ v(\mb{r}_{ij}) - \frac{q_iq_j}{||\mb{r}_{ij}||} \right]
\end{align}
is the Madelung constant which is independent of the configuration $X$, and depends only on $L$. Although its
contribution to the total energy is negligible in the limit $N_e \to \infty$, it is a standard term which is in general systematically included to accelerate the extrapolation to the thermodynamic limit.

For the homogeneous electron gas, electron density is usually expressed in terms of the Wigner-Seitz parameter $r_s=a/a_B=(4 \pi \rho_e a_B^3/3)^{-1/3}$, the ratio between the Bohr radius ($a_B=1$ in atomic units) and the mean interparticle distance $a=(4 \pi \rho_e/3)^{-1/3}$. Energies are given in Rydbergs. In the high density limit, $r_s \to 0$, kinetic energy dominates and the system approaches the ideal Fermi gas whereas at low densities kinetic energy quantum effects become less important and the electrons will eventually form a Wigner crystal to minimize the classical potential energy \cite{giuliani2005quantum}. 

\subsection*{Variational Monte Carlo}

Given that the Hamiltonian is a lower-bounded operator, the variational principle states that upper-bound estimates of the ground-state energy can be obtained by minimising the expectation value of the energy with respect to the parameters $\theta$ of an Ansatz $\psi(X; \theta)$  \cite{szabo2012modern}. The high dimensional integral required to compute the expectations can, in general, be performed with Monte Carlo integration. In this work, $M$ samples are drawn from $p(X;\theta) \propto |\psi(X;\theta)|^2$ in a Markov chain process via the Metropolis Hastings algorithm \cite{hastings1970monte}. For reviews of \gls{qmc} methods for solids see References~\cite{foulkes2001quantum, suzuki1993quantum, lester2009quantum}. 

The estimate of the energy is computed
\begin{align}
    \EX_{X\sim p(X;\theta)}\Big[E_L(X;\theta)\Big]
    \approx \frac{1}{M} \sum_{i=1}^M E_L(X_i;\theta)
\end{align} 
where the expectation is taken over the distribution $p(X;\theta)$ and this explicit notation is dropped from now on. $E_L(X; \theta)$, the local energy, is
\begin{align}
    E_L(X;\theta) = -&\frac{1}{2} \Big[ \nabla_X^2 \log \psi(X;\theta) \nonumber \\ +& \big(\nabla_X \log \psi(X;\theta) \big)^2 \Big] + V(X).
    \label{eq:local_energy}
\end{align} 
%
The derivatives of the expectation of the energy wrt the Ansatz parameters $\theta$ can be framed as the vector of derivatives $\nabla_\theta$ of a loss function  $\mathcal{L}(\theta)$ and are computed as 
\begin{align}
    \nabla_\theta \mathcal{L}(\theta) = \EX_X \Big[(E_L(X;\theta) - \EX_X&[E_L(X;\theta)]) \nonumber \\
    & \nabla_\theta \log \psi(X;\theta)  \Big].
    \label{eq:vmc_gradients}
\end{align}

\subsection*{Iterative Backflow Ansatz}

For fermions, the trial wave function, $\psi(X;\theta)=D(X;\theta) e^{-U(X;\theta)}$, is conveniently split into a generalized Jastrow factor, $U(X;\theta)$, symmetric with respect to any permutation in the particle labels $i$, and a manifestly antisymmetric from, $D(X;\theta)$, usually a Slater determinant composed of $N_e$ orbitals, $\psi_k(\cdot)$,
a set of plane waves for homogeneous systems, $\psi_{\mb{k}}(\mb{r})=\exp(i \mb{k} \cdot \mb{r})$. 

For a spin-independent Hamiltonian, we can further label the orbitals according to the spin polarisation, $\uparrow$/$\downarrow$, such that the Slater determinant factorizes, e.g.
$D(X)=[\det_{ij} \psi_{\mb{k}_j}(\mb{r}_i^\uparrow) ] \times [ \det_{ij} \psi_{\mb{k}_j}(\mb{r}_i^\downarrow)]$. Typically, the wave vectors   
used in the respective Slater determinant correspond to those of the ground state
wave function of the ideal Fermi gas. 

A standard two body Jastrow function writes $U_0(X)=\sum_{i<j} u_\theta^{0}(\mb{r}_{ij})$,\
 and backflow wave functions are obtained by the use of backflow coordinates $Q_0(X)=(\mb{q}_1^{0},\dots,\mb{q}_{N_e}^{0})$ with $\mb{q}^{0}_i=\mb{r}_i+\sum_{j \ne i} \mb{r}_{ij} \eta_\theta^{0} (\mb{r}_{ij})$ as arguments in the orbitals of the Slater determinant 
\begin{equation}
[\det_{ij} \psi_{\mb{k}_j}(\mb{q}_i^{0\uparrow}) ] \times [ \det_{ij} \psi_{\mb{k}_j}(\mb{q}_i^{0 \downarrow})] \exp[-\sum_{i<j} u_\theta^{0}(\mb{r}_{ij})]
\end{equation}

Higher order many-body correlations can then be systematically constructed by an iterative procedure $U_n(Q_{n-1})=\sum_{i<j} u_\theta^{n}(\mb{q}_{ij}^{n-1})$, and, analogously,
 new sets of backflow coordinates, $Q_n(X)$, from $\mb{q}^{n}_i=\mb{q}_i^{n-1}+\sum_{j \ne i} \mb{q}_{ij}^{n-1} \eta_\theta^{n} (\mb{q}_{ij}^{n-1})$
yielding the general form of
IB wave functions \cite{taddei2015iterative}
\begin{equation}
[\det_{ij} \psi_{\mb{k}_j}(\mb{q}_i^{n\uparrow}) ] \times [ \det_{ij} \psi_{\mb{k}_j}(\mb{q}_i^{n \downarrow})] \exp[- \sum_{n' \le  n} U_{n'}(Q_{n-1}) ]
\end{equation}
This structure not only captures many-body correlations in a compact way, but also allows for 
an efficient calculation of the wave function, its gradient and local energy, of order $N_e^3$
independent of the number $n$ of iterations.

Each iteration defines a new permutation equivariant structure, introducing a new set of parameters $\theta$ in the respective functions, $u_\theta^{n}(\cdot)$ and $\eta_\theta^{n}(\cdot)$,
possibly separated into spin-like and spin-unlike parts,
by use of a basis set expansion. 
Details of the implementation used here are given in Reference~\cite{BFStoner}.

\subsection*{WAP-net Ansatz}

The Ansatz developed here, dubbed \gls{wap}-net and represented in Figure~\ref{fig:wapnet}, is a progression on other implementations \cite{pfau2019ab, hermann2019deep, wilsonSimulationsStateoftheartFermionic2021} designed for molecular systems and is related to other periodic networks developed for bosonic systems \cite{pescia2021neural}. Key differences with previous work for molecular systems are embedding coordinates with periodic functions as in Reference~\cite{pescia2021neural}, and other changes to the Ansatz include splitting pairwise features by spin alignment and feeding backflow coordinates to the plane wave Hartree-Fock orbitals. In our implementation of the \gls{wap}-net we rescale the lengths so that the box side becomes $L=1$. The complete model is described below. 

The Ansatz comprises linear layers, second order differentiable non-linear activations, and Slater determinants. Before the Slater determinant, non-linear layers of functions operate on the electron coordinates ($\mb{r}_j$) and their displacements ($\mb{r}_{ij} = \mb{r}_j - \mb{r}_i$), to maintain indexes of intermediate variables such that they correspond to a coordinate or displacement from the inputs. Intermediate variables corresponding to coordinates and displacements flow through parts of the network called the single and pairwise streams, respectively. The permutation equivariance of these functions ensures that exchange of same spin electron coordinates results in the same permutation of the intermediate variables, resulting in the exchange of rows (or columns) of the Slater determinant, and flipping the amplitude sign, fulfilling the antisymmetry required for fermionic systems. 

Coordinates are first embedded by periodic functions to satisfy the periodicity of the system, including continuity of the function and derivatives across the boundary,
\begin{align}
    p(\mb{r}) = (\cos(2\pi \mb{r})&, \sin(2\pi \mb{r}), ..., \nonumber \\
     &...,  \cos(2n_p\pi \mb{r}), \sin(2n_p \pi \mb{r})),
\end{align}
where $n_p$ is the number of functions and the functions are applied elementwise. In the case of the pairwise streams a distance-like feature, $\norm{p_s(\mb{r}_{ij})}$, is computed and concatenated
\begin{equation}
    p_s(\mb{r}_{ij}) = \frac{1}{2}(\sin(\pi r^x_{ij}), \sin(\pi r^y_{ij}), \sin(\pi r^z_{ij}))
    \label{eq:sin_transform}.
\end{equation}
For the single streams, these periodic input features are concatenated with density fluctuations
\begin{equation}
    \rho_{{\bf k }}=  \Big( \sum_j\cos({\bf k}\cdot{\bf r}_j,  ), \sum_j\sin({\bf k}\cdot{\bf r}_j) \Big)
\end{equation}
given a set $\mathbb{K}$ of $n_k$ reciprocal lattice vectors $\mb{k}$. Overall, the variables in the zeroth layer are
\begin{align}
    \mb{h}^{0 \alpha}_i &= (\big\{\rho_{{\bf k}}: {\bf k} \in \mathbb{K}\big\}, p(\mb{r}_i)), \label{eq:fn_singleinputs} \\
    \mb{h}^{0\alpha\beta}_{ij} &= (p(\mb{r}_{ij}), \norm{p_s(\mb{r}_{ij})}), \label{eq:fn_pairwiseinputs}
\end{align}
Permutation equivariant functions then operate on data from both streams to compute single stream variables
\begin{align}
    \mb{f}^{l\alpha}_i &= \Bigg( \mb{h}^{l\alpha}, 
    \frac{1}{n_\uparrow}\sum_{ j \, \mathrm{if} \, \beta \neq \downarrow} \mb{h}^{l\alpha\beta}_{ij},
    \frac{1}{n_\downarrow}\sum_{j \, \mathrm{if} \,\beta \neq \uparrow} \mb{h}^{l\alpha\beta}_{ij}
    \Bigg) 
    \label{eq:fn_maxequivarianta} \\
    \mb{g}^l &= \Bigg(\frac{1}{n_\uparrow}\sum_{i \, \mathrm{if} \,\alpha \neq \downarrow} \mb{h}^{l\alpha}_i, \frac{1}{n_\downarrow}\sum_{i \, \mathrm{if} \,\alpha \neq \uparrow} \mb{h}^{l\alpha}_i
    \Bigg).
    \label{eq:fn_maxequivariantb}
\end{align}
Updates on the single and pairwise streams at layer $l$ are computed as
\begin{align}
    \mb{h}^{(l+1)\alpha}_i = &\cos\big(\mb{W}_{l}\mb{f}^{l\alpha}_i + \mb{Z}_{l}\mb{g}^{l} + \mb{b}_{l}\big) + \mb{h}^{l\alpha}_i  , \\
    \mb{h}^{(l+1)\alpha=\beta}_{ij} = &\cos\big( \mb{V}_{l} \mb{h}^{l\alpha=\beta}_{ij} + \mb{c}_{l} \big) + \mb{h}^{l\alpha=\beta}_{ij} \label{eq:pairwise_same}, \\ \mb{h}^{(l+1)\alpha\neq\beta}_{ij}= & \cos\big( \mb{Q}_{l} \mb{h}^{l\alpha\neq\beta}_{ij} + \mb{u}_{l} \big) + \mb{h}^{l\alpha\neq\beta}_{ij} \label{eq:pairwise_different},
\end{align}
for weights $\mb{W}_l$, $\mb{Z}_l$, $\mb{V}_l$, and $\mb{Q}_l$, and biases $\mb{b}_l$, $\mb{c}_l$, $\mb{u}_l$. Residual connections are added to all layers where $\mathrm{dim}(\mb{h}^l) = \dim(\mb{h}^{(l-1)})$. This implementation differs from previous similar works \cite{wilsonSimulationsStateoftheartFermionic2021, pfau2019ab, spencer2020better} as the pairwise stream layers are split into two spin-alignment dependent blocks: $\alpha=\beta$ and $\alpha\neq\beta$ corresponding to when the displacements are computed between electrons of the same and different spins, respectively. This method is motivated as correlations change depending on spin-alignment. 

\begin{figure}
    \centering
    \hspace*{-1cm} 
    \includegraphics[width=0.5\textwidth]{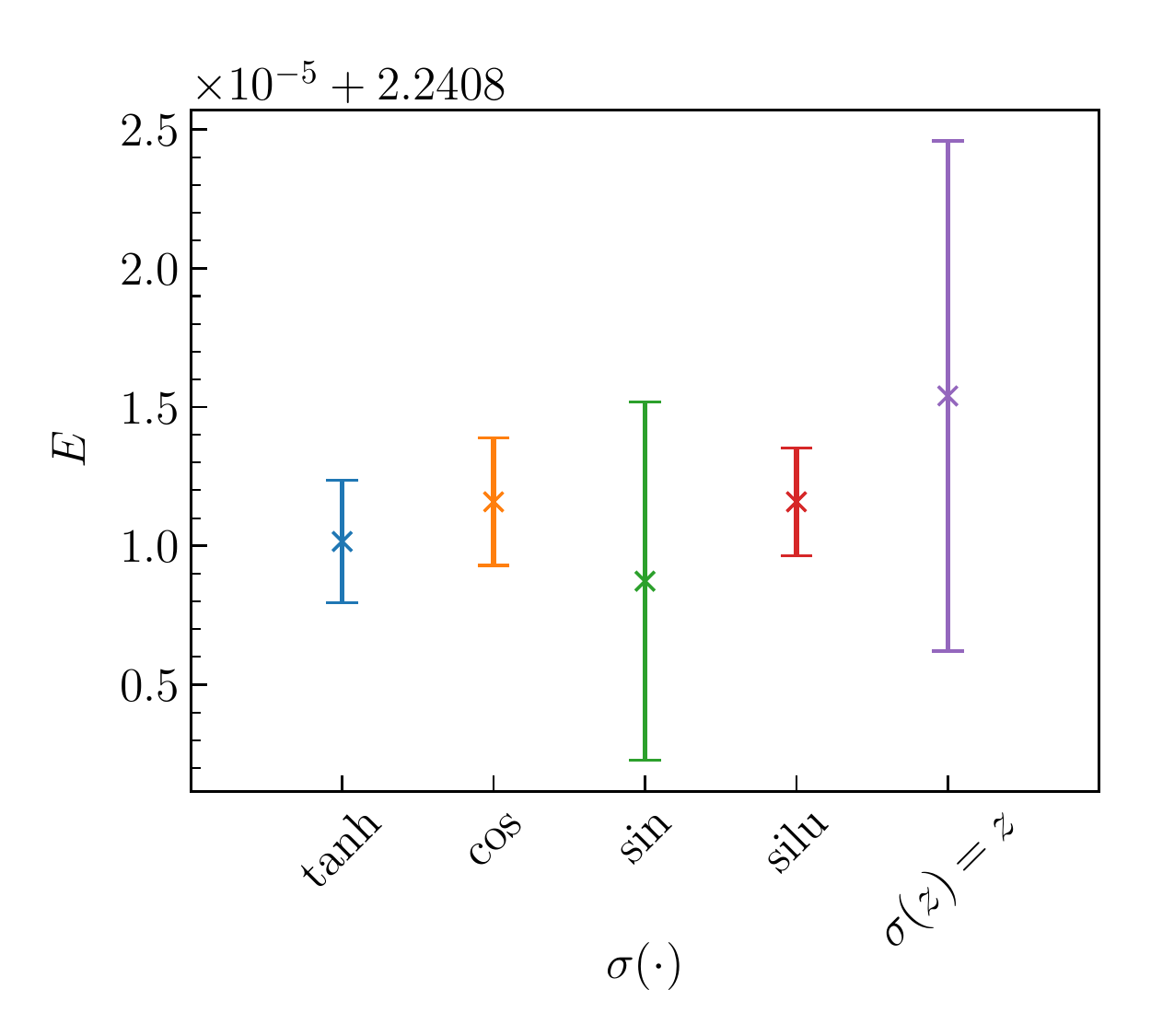}
    \caption{Comparison of different activation functions in the backflow layer, Equation~\ref{eq:backflow}, for $rs=1$ and $N=14$. Discussion on SILU can be found in Ref.\cite{elfwing2018sigmoid}. Mean and error bars represent the statisitics over a small ensemble of models trained with different seeds.}
    \label{fig:afcomparison}
\end{figure}

There are $n_l$ of these non-linear parameterised layers. The last layer is half the size of the previous layer and the outputs $\mb{f}^{L\alpha}_i$ and $\mb{g}^L$ are concatenated 
\begin{equation}
    \mb{h}^{L\alpha}_i = \mb{f}^{L\alpha}_i \mathbin\Vert \mb{g}^L,
\end{equation}
then split into spin dependent data blocks. These variables are first used to construct backflow coordinates as inputs to the Hartree-Fock orbtials 
\begin{equation}
    \mb{r}'^{\alpha}_j = \mb{r}^\alpha_j + \mathrm{tanh}(\mb{W}_L^{d\alpha} \mb{h}_j^{L\alpha} + \mb{b}^{d\alpha}_y),
    \label{eq:backflow}
\end{equation}
where $d$ indexes the Slater determinant where the backflow coordinates will be used, and a $\tanh$ activation was used to limit the outputs between -1 and 1. This gives good performance. Other activations were tested but the results showed no improvement and in the case of no activation function, the computed energies had higher variance and final energy, see Figure~\ref{fig:afcomparison}. Second, $\mb{h}_j^{L\alpha}$ are mapped to scalar orbital prefactors via a linear transformation $w^{d\alpha}_{i}(\cdot)$ giving Slater determinant orbitals
\begin{equation}
    \phi^{d\alpha}_{ij}(X) = 
     w^{d\alpha}_{i}(\mb{h}^{L\alpha}_j) \psi^H_i(\mb{r}'^\alpha_j)
     \label{eq:realspacehegorbitals},
\end{equation}
where $\psi^H_i(\mb{r}_j)$ are the Hartree-Fock plane wave orbitals 
\begin{equation}
    \psi^H_i(\mb{r}_j) =
    \begin{cases}
     \sin(\mb{k}_i \mb{r}_j) \quad & \mathrm{if} \, i \, \mathrm{is} \, \mathrm{even}, \\
     \cos(\mb{k}_i \mb{r}_j) \quad & \mathrm{if} \, i \, \mathrm{is} \, \mathrm{odd} \\
    \end{cases}
\end{equation}
where the set of $\mb{k}$ are ordered such that $\mb{k}_{i+1}$ is opposite to $\mb{k}_{i}$, i.e. $-\mb{k}_{i+1} = \mb{k}_{i}$, for even $i$ and $\mb{k}_1 = (0, 0, 0)$. We take the shortest
$N_\alpha$ k-points to form the Hartree-Fock orbitals for the $N_\alpha$ electrons of spin $\alpha$.

The Slater determinant of a spin becomes
\begin{equation}
    \det[\mb{\Phi}^{d\alpha}] = \begin{vmatrix}
\phi^{d\alpha}_{11}(X) & \hdots & \phi^{d\alpha}_{N_\alpha1}(X) \\ 
\vdots &  & \vdots \\ 
\phi^{d\alpha}_{1N_\alpha}(X) & \hdots & \phi^{d\alpha}_{N_\alpha N_\alpha}(X)
\end{vmatrix}.
\end{equation}
for $d$ defining the index.

The full wave function is written
\begin{align}
    \log |\psi(X)| = \log \Big|\sum_d \det[\mb{\Phi}^{d\uparrow}] \det[\mb{\Phi}^{d\downarrow}]  \Big|.
\end{align}
We have taken the log-absolute value of the amplitude in the implementation. While in general one can write $\log \psi = \log | \psi | + i \phi$, with the periodic boundary conditions adopted in this work the wave function is real. Therefore we can ignore the phase $\phi$ which is a constant wherever the modulus $| \psi |$ is non-zero.

The complete set of Ansatz hyperparameters is given in Table~\ref{tab:ansatz_hyperparameters}; these hyperparameters were chosen based on intuition from previous research \cite{pfau2019ab, wilsonSimulationsStateoftheartFermionic2021, spencer2020better} and sweeps over options (including for example alternate activations and hidden layer size). We tested different models and found best performance with the design outlined in this work, Figure~\ref{fig:model_evolution} portrays how the different features (splitting the pairwise stream by spin alignment and backflow coordinates) affected performance for a small demonstration example. Some early design choices behaved better than others, notably, the inclusion of a Jastrow factor $U(X)$ did not systematically improve the performance, similar to atomic and molecular
systems \cite{pfau2019ab}. This suggests that the model is able to efficiently describe the electron-electron cusp conditions \cite{foulkes2001quantum} at low $r_s$ and the strong curvature of the correlation hole \cite{foulkes2001quantum} at large $r_s$. The model and algorithm were implemented in Jax \cite{jax2018github} and run on RTX 3090 GPUs.

\begin{table*}
\centering
\noindent\begin{tabularx}{0.6\textwidth}{>{\hsize=.5\hsize}X >{\hsize=2\hsize}X >{\hsize=.5\hsize}X}
\textbf{Symbol} & \textbf{Description} & \textbf{Value} \\ \hline
N/A & Single stream hidden units & 128 \\ \hline
N/A & Pairwise stream hidden units & 32 \\ \hline
$n_l$ & Number of layers & 3 \\ \hline
$n_p$ & Number of periodic functions & $5$ \\ \hline
$n_d$ & Number of determinants & \{1, 8\} \\ \hline
$n_k$ & Number of k-points used in the input & 19 \\ \hline
\end{tabularx}
\caption{Hyperparameters used to define the model. }
\label{tab:ansatz_hyperparameters}
\end{table*}

\subsection*{Optimisation}

Before parameters are optimised via \gls{vmc}, the Ansatz is first pretrained on Hartree-Fock orbitals using methods described in other References \cite{wilsonSimulationsStateoftheartFermionic2021, pfau2019ab}. At a high level, Ansatz orbitals in the Slater determinants are fit to Hartree-Fock orbitals in a supervised way: A loss is computed as the squared difference between them. This helps the Ansatz to start closer to the ground-state, to avoid being stranded in local minima, and not to diverge during early training due to large gradients. 

Additionally, during \gls{vmc} optimisation, parameter gradients computed from Equation~\eqref{eq:vmc_gradients} are transformed to approximate natural gradients, using a \gls{kfac} \cite{grosse2016kronecker, martens2015optimizing, ba2016distributed} approximation to the Fisher Information Matrix, 
\begin{equation}
    \boldsymbol{\delta}_l = \EX_{\mb{a}_l}[ \mb{a}_l \mb{a}_l^T ] \nabla_\theta L(\theta) \EX_{\mb{s}_l}[ \mb{s}_l \mb{s}_l^T ],
\end{equation}
where $\mb{a}$ and $\mb{s}$ are the activations and sensitivities \cite{martens2015optimizing}, respectively, and $l$ is the layer index. Though correct, this equation ignores a complete description of the algorithm used to compute the approximate natural gradients, which requires damping, constraining of the norm of the transformed gradients, and smoothing approximations to the covariance matrices. Particularly, damping and the covariance matrices are computed via methods developed for convolutional layers \cite{grosse2016kronecker}. Norm constraint and other details of the method are described in the Appendix of Reference~\cite{wilsonSimulationsStateoftheartFermionic2021}, more background on these optimisation routines can be found in References~\cite{grosse2016kronecker, martens2015optimizing}, and a list of optimisation hyperparameters can be found in Table~\ref{tab:kfac_hyperparameters}.

Finally, similar to stochastic reconfiguration \cite{Sandro2001}, the optimal step size in a natural
gradient descent scales roughly with the inverse of the energy. In order to capture
variations between $r_s=1$ and $r_s=100$,
the gradients are heuristically scaled depending on the density parameter of the system
\begin{equation}
    \boldsymbol{\delta}'_l = r_s^{1+r_s/100} \boldsymbol{\delta}_l
\end{equation}
This helps
to amplify the gradients when the variance of the energy is small, particularly  for systems with large $r_s$. This heurisitic is essential to achieving state-of-the-art accuracy for these systems for both spin-polarised ($N_\uparrow=N_e$) and paramagnetic ($N_\uparrow=N_\downarrow=N_e/2$) systems.

\begin{table}
\centering
\noindent\begin{tabularx}{0.45\textwidth}{>{\hsize=1.9\hsize}X >{\hsize=0.7\hsize}X >{\hsize=0.7\hsize}X >{\hsize=0.7\hsize}X}
\textbf{Description} & \textbf{Value} & \textbf{Decay} & \textbf{Floor} \\ \hline
Learning rate & $1\times10^{-3}$ & $1\times10^{-4}$ & $1\times10^{-4}$\\ \hline
Damping & $1\times10^{-4}$ & $1\times10^{-2}$ & $1\times10^{-6}$\\ \hline
Norm constraint & $1\times10^{-4}$ & $1\times10^{-4}$ & $1\times10^{-6}$ \\ \hline
Number of walkers & 2048 & N/A & N/A \\ \hline
Number of iterations & $1\times10^5$ & N/A & N/A \\ \hline
\end{tabularx}
\caption{Table of KFAC hyperparameters.}
\label{tab:kfac_hyperparameters}
\end{table}

\section{Results}\label{sec:results}
\subsection{Ground state energy}\label{subsec:energy}
\begin{figure}
    \centering
    \includegraphics[width=0.5\textwidth]{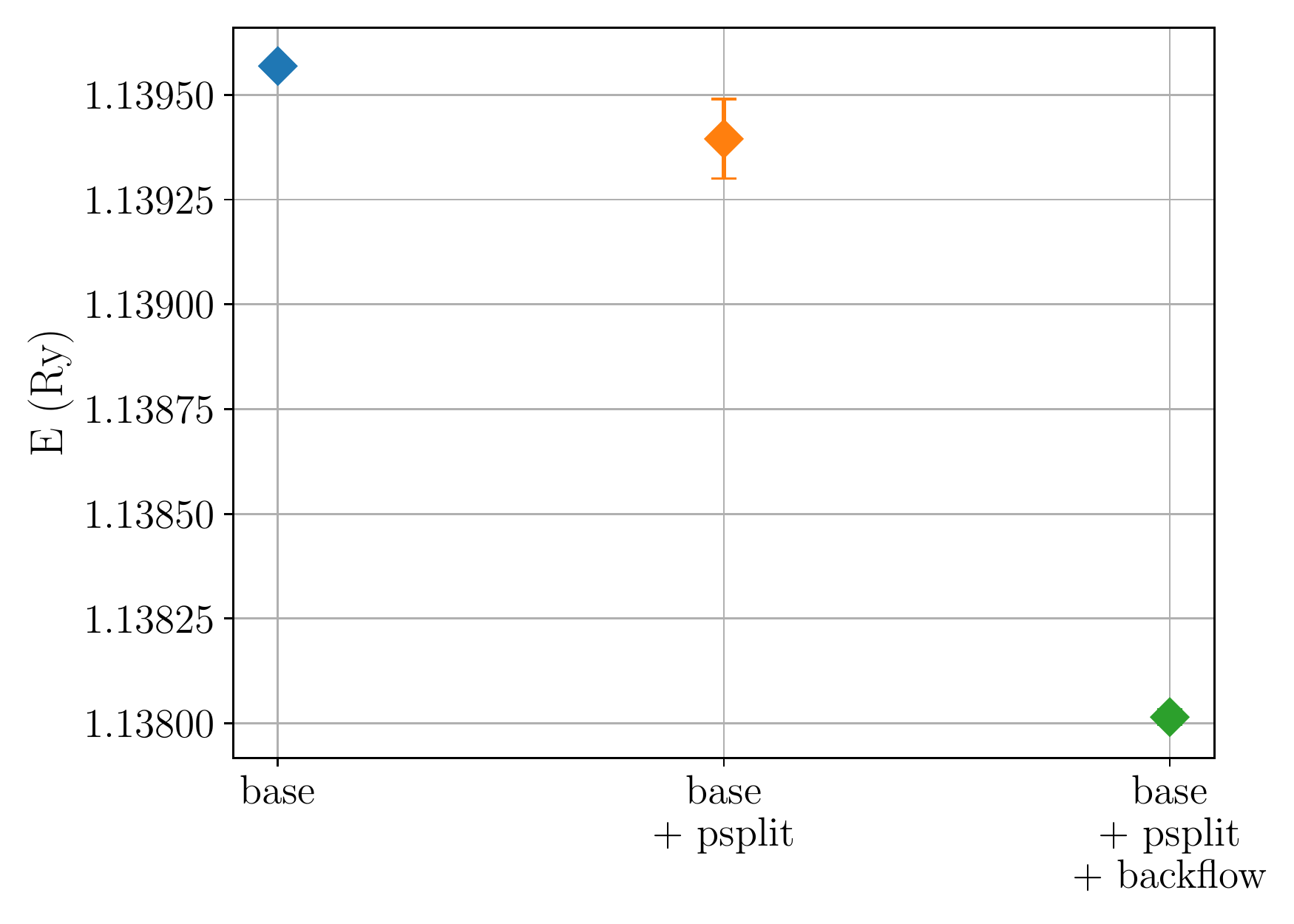}
    \caption{Plot showing the change in performance for a small model on a paramagnetic ($N_e=14$) system over $3\times10^4$ training iterations. psplit indicates the method of splitting the pairwise streams by spin alignment (Equations~\eqref{eq:pairwise_same}-\eqref{eq:pairwise_different}) and backflow the adjustment to the Hartree-Fock orbital inputs (Equation~\eqref{eq:backflow}). The error bars (only visible for base+psplit) are average over the 2 runs for the standard error on the mean of the energy calculation, computed with 1000 batches of 512 walkers.}
    \label{fig:model_evolution}
\end{figure}

\begin{figure*}
    \centering
    \includegraphics[width=0.9\textwidth]{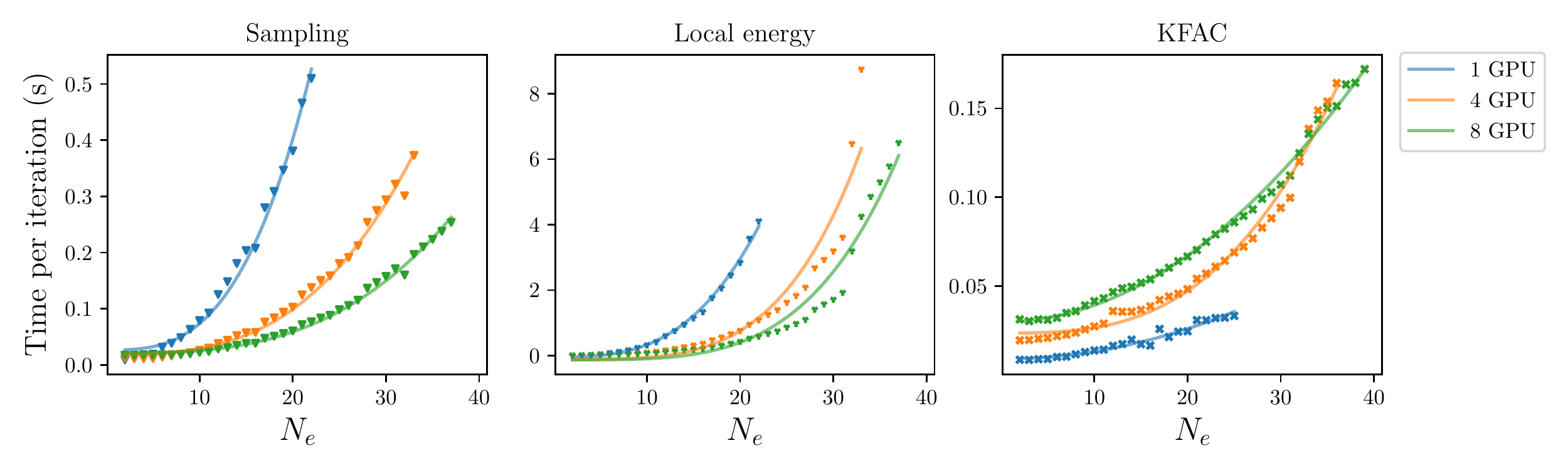}
    \caption{Plots containing timing data for subroutines (Sampling - left, Local energy - middle, and KFAC - right) for spin-polarised systems as a function of the number of electrons $N_e$. Each point represents the average of 1000 executions of the subroutine and each plot contains three lines representing the subroutines run with 1 (blue), 4 (orange), and 8 (green) GPUs. Fit lines are the best fit (the smallest mean squared residual) from a selection of 3 polynomials of the form $f(N_e) = a N_e^p + c$, where a and c are parameters and $p \in \{2, 3, 4\}$. In most cases, the Sampling routine scales as $\mathcal{O}(N_e^3)$, the Local energy as $\mathcal{O}(N_e^4)$, and KFAC as $\mathcal{O}(N_e^2)$. The only cases where this didn't hold were for the local energy subroutine in the spin-polarised system with 1 GPU, finding a cubic fit, however the difference between the mean residuals of the fit between the cubic and quartic cases was small (within ~2\% mean residual).}
    \label{fig:timing_data}
\end{figure*}

In order to illustrate the accuracy of \gls{wap}-net, we have computed VMC and fixed-node DMC energies based on the iterative backflow (IB) network wave function. Our IB results of Table~\ref{tab:results} outperform the corresponding best single-determinant BF-VMC and BF-DMC values previously published ($N=19$, $r_s=1$: Table 5 in the Supplementary Material of \cite{ruggeri2018correlation}, $N=14$, $r_s=\{1,2,5,10,20,50\}$ in Reference~\cite{Transcorr}). Comparing to extrapolated transcorrelated FCIQMC \cite{Transcorr} available for $N=14$ at high densities, $r_s \le 5$, we estimate the residual fixed-node error to be around $0.7$ mRy or below in this region, becoming more accurate with decreasing density. For the spin-polarised system, IB-DMC values are expected to be in general closer to the exact ground-state values, e.g. around $0.2$ mRy for $N=19$, $r_s=1$ \cite{ruggeri2018correlation}. Therefore, our baseline IB results provide a challenging benchmark for \gls{wap}-net over the full density region of $r_s=1$ to $r_s=100$ for the spin-polarised and paramagnetic system.

During the design phase of \gls{wap}-net many different model features and implementations were tested. Representative results we include here from those tests are shown in Figure~\ref{fig:model_evolution} highlighting the hierarchy of performance observed whilst developing the model. Base defines a model with periodic input features and permutation equivariant layers. Psplit adds split pairwise layers (Equations~\eqref{eq:pairwise_same} and~\eqref{eq:pairwise_different}), and backflow indicates channeling backflow coordinates from the permutation equivariant layers output to the Hartree-Fock orbitals in a backflow-like way, showing in this small example a clear improvement in performance as these features are added. 

The following results refer to the \gls{wap}-net with model and \gls{kfac} hyperparameters stated in Tables~\ref{tab:ansatz_hyperparameters} and \ref{tab:kfac_hyperparameters}, respectively, for example the number of training iterations was $1\times10^5$ for each experiment. For the comparative methods (\gls{wap}-net with $n_d=1$ and the \gls{ib} baseline, which was run with a single Slater determinant, second and fourth column in Table~\ref{tab:results}, respectively, counting the density parameter $r_s$ as the first column), we find that our approach improves upon previous \gls{vmc} results in all cases $r_s \in \{1, 2, 5, 10, 20, 50, 100\}$ for systems $N_e \in \{7, 14\}$, and most $r_s \in \{1, 2, 5, 10, 20\}$ for the spin-polarised $N_e = 19$ case. Additionally, in all cases, other than $r_s=2$, for the spin-polarised $N_e=7$ systems, \gls{wap}-net improved upon the \gls{ib} wave function with \gls{dmc}. In some cases, $r_s \in \{1, 2\}$ for  $N_e=14$ and $r_s=1$ for $N_e=19$, respectively, \gls{wap}-net improved upon the \gls{ib} wave function with \gls{dmc}.

We additionally ran networks with $n_d=8$ and found no consistent improvement over the single determinant case. The results were marginally improved in the cases $N_e=7$ $r_s \in \{1, 2, 5, 10, 20\}$, $N_e=14$ $r_s \in \{5, 10\}$, and $N_e=19$ $r_s \in \{1, 2, 5, 10, 50\}$. 

In order to understand the scaling of \gls{wap}-net with respect to system size we computed the average time taken for 1 iteration of each of the subroutines (Sampling, Local energy, and \gls{kfac}), as seen in Figure~\ref{fig:timing_data}. Each set of timing data was taken with 1, 4, and 8 GPUs. Polynomials of form $f(N_e) = aN_e^p + c$ were fit to the data where $a$ and $c$ were adjustable parameters and $p\in \{2, 3, 4\}$. Lines were fit to each of the datasets and the best fit (as measured by the minimum squared residual) was chosen. The cost of the subroutines scales as expected: $\mathcal{O}(N_e^3)$ for Sampling, which is dominated by the determinant computation (known to scale as $\mathcal{O}(N_e^3)$); $\mathcal{O}(N_e^4)$ for the Local Energy computation (which in some cases was close to $\mathcal{O}(N_e^3)$), and can be understood in terms of the scaling as $N_e$ evaluations of the determinant; and finally $\mathcal{O}(N_e^2)$ in the case of \gls{kfac}. There is a small discontinuity in the Local Energy data around $N_e=32$ for 4 and 8 GPUs. This might be explained by reaching some threshold of data in the GPUs, for example if operations need to be queued as the number of variables increase.
\begin{table*}
\centering
\noindent\begin{tabularx}{0.9\textwidth}{|>{\hsize=.4\hsize}X | >{\hsize=1.15\hsize}X >{\hsize=1.15\hsize}X >{\hsize=1.15\hsize}X >{\hsize=1.15\hsize}X |}
\hiderowcolors \multicolumn{5}{c}{$N_e = 7$ (spin-polarised)} \\
\multicolumn{5}{c}{} \\
\multicolumn{1}{c}{$r_s$} & \specialcell{\textbf{WAP-Net} \\ ($n_d=1$)} & \specialcell{\textbf{WAP-Net} \\ ($n_d=8$)} & $E^\mathrm{IB}_\mathrm{VMC}$ & \multicolumn{1}{l}{$E^\mathrm{IB}_\mathrm{DMC}$} \\ \hline
1.0  &2.240785(7)       &2.240715(7)     & 2.24084(2)  &  2.24080(1) \\ \hline
2.0  &0.221766(5)       &0.221700(2)     & 0.221803(3)  & 0.221758(4) \\ \hline
5.0  &-0.132825(2)      &-0.132845(1)    & -0.132773(1) &  -0.132811(1) \\ \hline
10.0 	 &-0.1062797(5) 	 &-0.1062872(5)       & -0.1062376(6) &  -0.106276(1) \\ \hline
20.0 	 &-0.0645378(2) 	 &-0.0645379(3)      &-0.0644823(3) &  -0.064533(4)  \\ \hline
50.0 	 &-0.02932178(3)	 &-0.02932161(4)     &-0.0292957(1) &   -0.0293208(2) \\ \hline
100.0	 &-0.01549309(2)	 &-0.01549304(1)      &-0.01547628(7) & -0.0154925(1)  \\ \hline
\multicolumn{5}{c}{} \\
\hiderowcolors  \multicolumn{5}{c}{$N_e = 14$ (paramagnetic)} \\
\multicolumn{5}{c}{} \\
\multicolumn{1}{c}{$r_s$} & \specialcell{\textbf{WAP-Net} \\ ($n_d=1$)} & \specialcell{\textbf{WAP-Net} \\ ($n_d=8$)} & $E^\mathrm{IB}_\mathrm{VMC}$ & \multicolumn{1}{l}{$E^\mathrm{IB}_\mathrm{DMC}$} \\ \hline
1.0  &1.137912(9)          &1.13793(1)   & 1.13832(1)          & 1.13795(1)   \\ \hline
2.0  &-0.016666(5)         &-0.016662(5)  &   -0.016408(5)      &-0.01665(1)   \\ \hline
5.0  &-0.159669(1)         &-0.159672(1)    & -0.159524(1)     & -0.159684(3)   \\ \hline
10.0 	 &-0.1104049(6)   	 &-0.1104076(6)   & -0.1103229(7)     & -0.110436(1)   \\ \hline
20.0 	 &-0.0648883(3)   	 &-0.0648868(2)     & -0.0648444(2)    &  -0.0649157(5)   \\ \hline
50.0 	 &-0.02924445(6)  	 &-0.02924422(7)     & -0.02922920(2)   & -0.0292573(2)   \\ \hline
100.0	 &-0.01546048(3)  	 &-0.01545996(3)         & -0.01545525(6)      & -0.01546859(5)   \\ \hline
\multicolumn{5}{c}{} \\
\hiderowcolors  \multicolumn{5}{c}{$N_e = 19$ (spin-polarised)} \\
\multicolumn{5}{c}{} \\
\multicolumn{1}{c}{$r_s$} & \specialcell{\textbf{WAP-Net} \\ ($n_d=1$)} & \specialcell{\textbf{WAP-Net} \\ ($n_d=8$)} & $E^\mathrm{IB}_\mathrm{VMC}$ & \multicolumn{1}{l}{$E^\mathrm{IB}_\mathrm{DMC}$} \\ \hline
1.0   &2.092495(5)       &2.092483(5)   &  2.092544(7)       &   2.092450(1)  \\ \hline
2.0   &0.192608(1)       &0.192606(2)   & 0.192637(3)  &  0.192605(3) \\ \hline
5.0   &-0.1345001(7)     &-0.1345021(5)&  -0.134480(1)   & -0.134508(1)  \\ \hline
10.0 	 &-0.1057197(2)  	 &-0.1057207(2)   &  -0.1057090(6)     & -0.1057304(6) \\ \hline
20.0 	 &-0.0640219(1)  	 &-0.0640164(2)   & -0.0640200(2)   &  -0.0640333(2) \\ \hline
50.0 	 &-0.02912995(5) 	 &-0.02913142(4)  & -0.02913147(5)   & -0.02913839(8)   \\ \hline
100.0	 &-0.01543521(2) 	 &-0.01542724(3)    &-0.01543798(2)    & -0.01544285(2)  \\ \hline

\end{tabularx}
\caption{Results obtained by the simulations described in this paper. The reported values are the energies per particle in Rydbergs and the titles spin-polarised ($N_\uparrow=N_e$) and paramagnetic ($N_\uparrow=N_\downarrow=N_e/2$) describe the systems. Errors are the standard error on the mean represented as bracketed numbers with the same precision as the last digit, for example 2.13(2) is equivalent to $2.13 \pm 0.02$, where 0.02 is the standard error on the mean of 2.13. For \gls{wap}-net, the energies and errors are computed from 1000 batches of 2048 walkers. \label{tab:results}}
\end{table*}
\subsection{Discussion of performance}\label{subsec:discussion}
Whereas our network achieved an overall excellent accuracy compared to the \gls{ib} baseline, the quality deteriorates for larger $r_s$ and larger systems. This trend might be somewhat expected. The network hyperparameters are kept the same across all systems meaning the setup chosen would need to be extended: Adding more or larger hidden layers, more walkers (to approximate the natural gradients better) or training for more iterations to tackle more complex systems to the same relative performance as the less complex systems (fewer electrons and smaller $r_s$).

Having that in mind, the energy gain with respect to \gls{ib}-VMC calculations is comparable to those from \gls{ib}-DMC for $N=7$ and $N=19$ providing a rough demonstration of size consistency of the \gls{wap}-net Ansatz. However, further tests on larger systems are needed to fully confirm size-consistency in practice.

In contrast to previous networks on atomic and molecular systems \cite{pfau2019ab, hermann2019deep}, we do not observe a consistent improvement increasing the number of Slater determinants in the wave function. This may simply be a consequence of the translational symmetry when considering extended systems and working with closed shell situations, as multi-determinant wave functions do not considerably improve the ground-state energies of noble gas atoms, e.g. Ne \cite{pfau2019ab,brown07}.

Our results indicate quadratic scaling of the the \gls{kfac} subroutine. We expected linear scaling in the number of electrons as adding additional electrons only increases the number of layers for which a \gls{kfac} step is performed (computation of the covariances and their inversion). It is possible to see this trend in the 1 GPU case and observe that adding additional GPUs adds overhead for moving data between them, potentially explaining the difference between expected and observed scaling. Nevertheless, this subroutine is relatively negligible (in terms of the absolute time) for a complete iteration of training. 

In testing and in other works \cite{pfau2019ab, wilsonSimulationsStateoftheartFermionic2021, spencer2020better}, \gls{kfac} shows clear advantages in optimisation over more standard techniques such as Adam \cite{hermann2019deep, scherbelaSolvingElectronicSchrodinger2021}. However, the algorithm requires careful balancing of the hyperparameters (learning rate, damping, and norm constraint) to achieve optimal performance, which consumes time in development and tweaking of model design. Alternate methods such as conjugate gradient, which approximate the Fisher Information Matrix directly \cite{neuscammanOptimizingLargeParameter2012} and has been demonstrated in other works \cite{kingmaAdamMethodStochastic2014, gao_pesnet_2022}, may improve results or require less heuristics. 
\subsection{Single-particle density matrix}\label{subsec:obs}

So far, our discussion was naturally focused on the calculation of the energy expectation value, 
intrinsically connected to all variational approaches. Parametrizations
of the HEG correlation energy from QMC calculations, provide
the input of practically all Kohn-Sham density functionals (DFT) based on the local densiy approximation (LDA). Although our WAP-net and IB results provide the basis to reduce the fixed-node
error of previous calculations based on Slater-Jastrow or simple backflow wave function,
current DFT functionals
are more likely affected by corrections to LDA than by these comparably small changes in the HEG energies.
On the other hand, improvements of the underlying wave function may be more relevant to correlation functions which encode direct physical insight.

One of the physical observables where electron-electron interactions
are directly visible is the electronic momentum distribution. Whereas electrons only occupy
momentum states up to the Fermi surface in an ideal Fermi gas, electronic correlations
also involve the occupation of states above the Fermi surface. This leads to a reduction
of the size of the jump of the momentum distribution at the Fermi surface, the so-called
renormalization factor $Z$. Values of $Z$ for the HEG are relevant both qualitatively, as the defining feature of the Fermi liquid paradigm, and quantitatively, to explain experimental 
measurements of the momentum distribution in solid Na \cite{Na} and Li \cite{Li}, 
corrected by band structure \cite{ComptGw} and core electron effects \cite{Licore}.

In the following we will focus on the reduced single
particle density matrix
\begin{equation}
    g_1(r) =
\EX_{X \sim p(X)}
\left[ \frac{
\psi({\bf r}_1+{\bf r},{\bf r}_2,\dots {\bf r}_N) }{\psi({\bf r}_1,{\bf r}_2,\dots {\bf r}_N)}\right]  
\label{Eq:g_1}
\end{equation}
where $p(X)$ is proportional to $|\psi(X)|^2$.
From $g_1(r)$ the momentum distribution can then be obtained by Fourier transform.

Having an explicitly parameterised wave function $\psi(X)$, it is straightforward to compute
$g_1(r)$ within VMC \cite{McMillan}. The DMC calculation, instead, is more involved.
Within DMC the trial wave function, $\psi(X)$, is stochastically projected
to its fixed-node ground state, $\psi_0(X)$,
resulting in lower energy values. However, 
the configurations $X$ are now sampled
according to the weight $p_{DMC} \sim |\psi_0(X) \psi(X)|$.
For a general operator $A$, for which $\psi_0$ is not an eigenfunction,
the calculation of the expectation value $\langle \psi_0  |A | \psi_0 \rangle$
based on the mixed distribution $p_{DMC}$ typically introduces a 
large variance \cite{LiuKalosChester1974}.     
Alternatively, the mixed estimator bias can be avoided by reptation Monte Carlo (RMC) methods \cite{rept}; however, the application of RMC to off-diagonal properties like $g_1(r)$ is rather elaborated \cite{momk3D}.
Instead, we follow here the common DMC practice to assume $\psi$ sufficiently close
to $\psi_0$, so that
\begin{eqnarray}
\langle \psi_0  |A | \psi_0 \rangle
& \approx & \langle \psi |A | \psi \rangle \\
& + & \left[ \langle  \psi_0 | - \langle \psi | \right]
A |\psi \rangle + \langle \psi | A  \left[ | \psi_0 \rangle  - |\psi \rangle \right],\nonumber
\label{eq:mix}
\end{eqnarray}
leading to the so-called extrapolated estimator  \cite{extrapol} 
\begin{equation}
A_{ext} \equiv 2 \EX_{X \sim p_{DMC}(X)}[A] - \EX_{X \sim p(X)}[A].
\end{equation}

\begin{figure}
\includegraphics[width=0.5\textwidth]{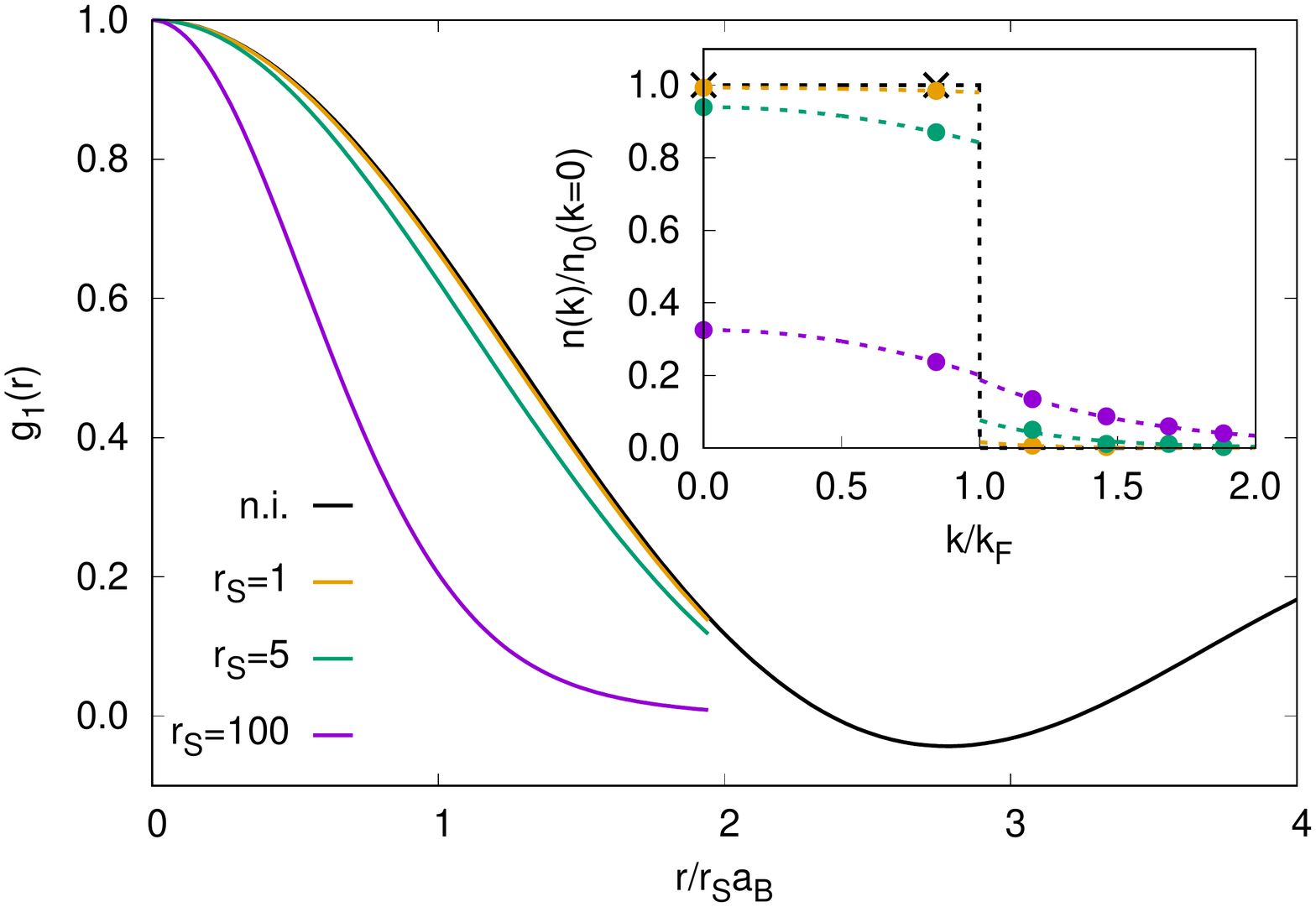}
    \caption{The reduced single particle density matrix $g_1(r)$
    of the paramagnetic HEG. The coloured lines are extrapolated
    estimates calculated with IB wave functions for $r_S=1$, $5$ and $100$ at $N_e=14$. The black line is the analytic result for the non-interacting Fermi gas at the same system size. The inset
    shows the corresponding momentum distributions (dotted lines are a guide to the eye)}
    \label{fig:g1andnk}
\end{figure}

Although the extrapolated estimator can reduce the bias and eventually recover the exact
expectation value of local operators (see Ref. \cite{fse} for an application to long-range properties
of the pair correlation function), this is not guaranteed for non-local observables such as $g_1(r)$: key physical properties such as the presence of a condensate for Bose systems \cite{ReattoChester1967} or a finite $Z$ for Fermi systems \cite{SaverioMomkHe} are hard-coded in the analytic form of standard wave functions, and they cannot be modified by either the mixed or the extrapolated DMC estimators. In practice, the Fermi liquid nature of the HEG is guessed in advance and maintained through the IB optimization and subsequent DMC projection, and the study of competing phases typically requires generalizations of the wave function specific to the targeted state of the system.

The WAP-net ansatz, on the contrary, is essentially 
agnostic, because it addresses individually each of the 
$N_\alpha^2$ orbitals of Equation~\ref{eq:realspacehegorbitals} through the many-body 
prefactors $w_i^{d\alpha}({\bf h}_j^{L\alpha})$.
This has the potential to alter the Fermi liquid nature expressed by filling the $N_a$ lowest-energy plane waves $\psi_i^H({\bf r}_j'^{\alpha})$.

We now present our results for $g_1(r)$ obtained with SJ and IB wave functions, and then discuss the insight we gain using the WAP-net ansatz. Here, SJ denotes 
a wave function based on a Slater determinant and
Jastrow function using only bare electron coordinates.

\begin{figure}
\includegraphics[origin=c,width=0.5\textwidth]{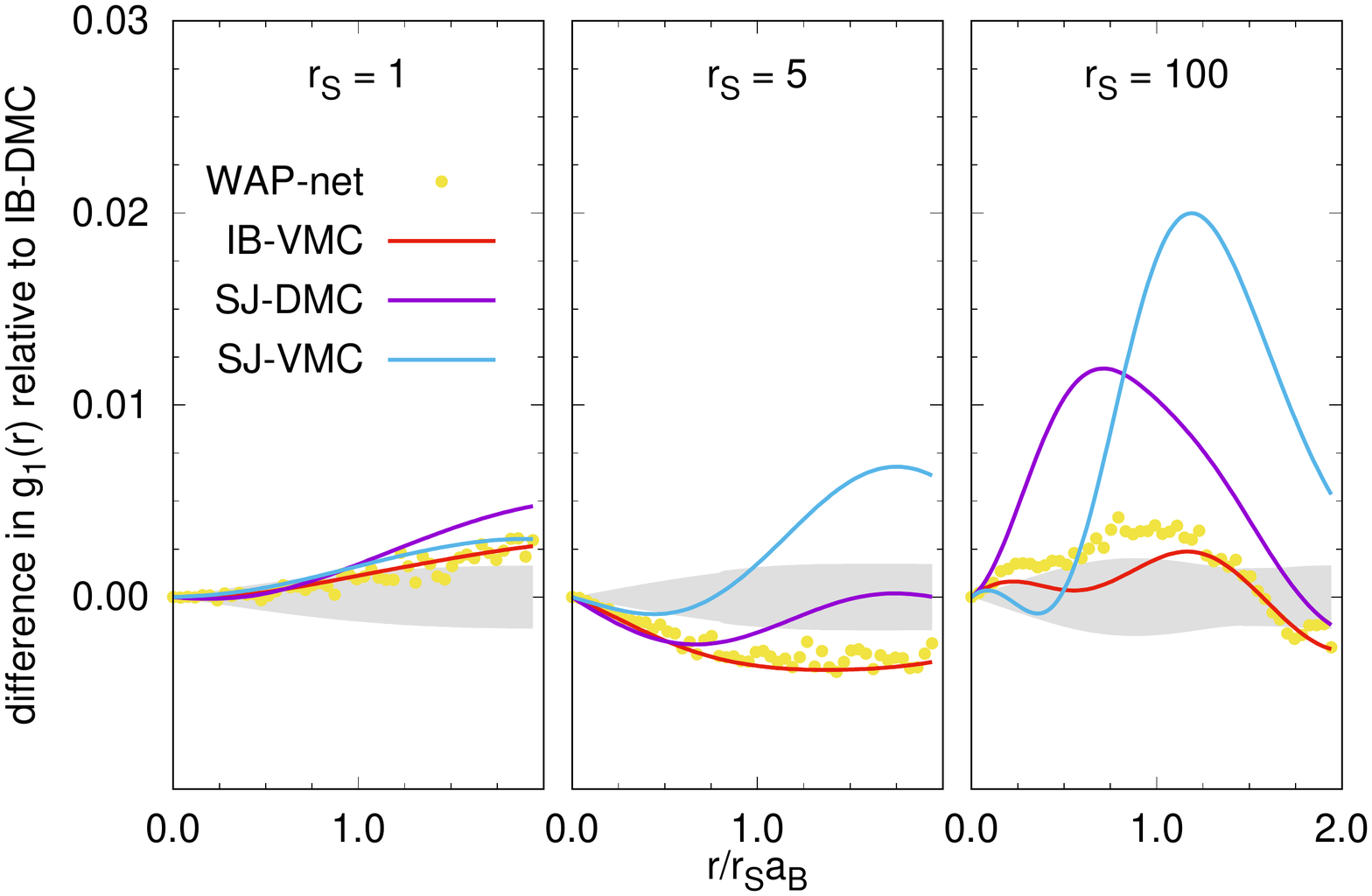}
    \caption{The difference between various calculations of $g_1(r)$ and the
    IB extrapolated estimate IB-DMC of Fig. \ref{fig:g1andnk}. The shaded area indicates the statistical uncertainty of IB-DMC. The
    WAP-net result is very close to the IB variational estimate IB-VMC.}
    \label{fig:g1rel}
\end{figure}

Figure~\ref{fig:g1andnk} illustrates the reduced single particle density matrix and the
resulting momentum distribution for the unpolarized $N=14$ electron system 
at three different densities, $r_s=1$, $5$, and $100$ using the extrapolated estimator with IB wave functions.
Increasing $r_s$
(decreasing density),
 $g_1(r)$ progressively departs from the ideal gas
curve. In the momentum distribution, shown in the inset, these deviations correspond to a reduction of the jump, $Z$, at the Fermi wave vector.

Differences between $g_1(r)$ obtained from optimized SJ, IB, and WAP-net wave functions relative to those of
extrapolated IB-DMC are shown in Figure~(\ref{fig:g1rel}).
Under the plausible assumption that the IB-DMC values are more accurate, the SJ-VMC and SJ-DMC results indicate that the DMC extrapolated estimates are not uniformly better than  VMC across the range of density and distance
explored. This may cast doubt on the full reliability of the correction from the IB-VMC to the IB-DMC results.

The WAP-net ansatz offers a completely independent route to the calculation of $g_1(r)$. Figure \ref{fig:g1rel} shows that the 

WAP-net values of $g_1(r)$ are much closer to IB-VMC than IB-DMC. While this may seem disappointing at first, there are good reasons to prefer the WAP-net to the IB-DMC results. The energies listed in Table \ref{tab:results} imply that the WAP-net wave function $\psi$ is nearly as good or better, depending on $r_s$, than the fixed-node IB-DMC projected eigenstate $\psi_0$. The WAP-net estimate of $g_1(r)$ is better than IB-DMC because it uses an approximation to the true ground state of similar quality and it does not make the further assumption of Equation~(\ref{eq:mix}).

Having better control on $g_1(r)$, we will now
conclude on assessing the accuracy of existing QMC
of $Z$ for the HEG  \cite{momk3D},
addressing possible sources of discrepancy with experiment \cite{Na,Li}, as well as with recent Diagrammatic Monte Carlo calculations \cite{Dia}.
The momentum distribution as well as the single
particle density matrix
shown in Figure~\ref{fig:g1andnk}) suffer from strong finite size
effects, and thermodynamic limit extrapolation is beyond the scope of the present article. 
Nevertheless, noting that leading order finite-size corrections are linear in $g_1(r)$ \cite{momk3D}, we
can study the systematic error due to quality of the wave function,
which roughly decouples from that of the system size.

From Figure~\ref{fig:g1rel} we see that IB-VMC 
agrees well with WAP-net at all three densities
whereas changes between SJ-VMC and WAP-net increase
with increasing $r_s$.
The corresponding variations in the momentum distribution would indicate 
a lowering of $Z$
compared to optimized SJ-VMC 
of $\sim 0.03$ for $r_s=5$ 
and $\sim 0.04$ for $r_s=100$, whereas at $r_s=1$  differences
are below our resolution.
Calculations done in Ref.~\cite{momk3D} are
based on analytical Slater-Jastrow and (non-iterated)
backflow wave functions. The quality of BF-RMC
in \cite{momk3D} is therefore expected to lie between our optimized SJ-VMC and IB-VMC/WAP-net ones.
Thus, the systematic error due to the quality of the wave
function used in \cite{momk3D} is less than the
error quoted, whereas $Z$ may lower by at most $3$\% at $r_s=5$. Although small, such  a shift may be relevant for
improving comparison with experiments \cite{Na,Li},
but would also place them slightly below those of
diagrammatic Monte Carlo calculations \cite{Dia}.

\section{Conclusion}\label{sec:conclusion}
In this work we have presented a neural network Ansatz, \gls{wap}-net, for approximating the many-electron ground state of the Homogeneous Electron Gas, extending previous work on atoms and molecules to periodic  electronic systems. Using the \gls{wap}-net Ansatz with Variational Monte Carlo methods, we have obtained accuracies comparable to or beyond our best Diffusion Monte Carlo calculations based on Iteractive Backflow wave functions for spin-polarised and paramagnetic systems over a broad density region.  The comparison of results for polarised systems from $N_e=7$ to $N_e=19$ shows that the Ansatz is size consistent (at least in the single determinant case, $n_d=1$). Larger systems are within reach, such that thermodynamic limit values can be properly addressed based on finite size extrapolations \cite{fse}. This method can be extended to complex wave functions with more general twisted boundary conditions \cite{lin2001twist} to accelerate the thermodynamic limit extrapolation.

Applying the \gls{wap}-net Ansatz to the Homogeneous Electron Gas, we focused on the description of intrinsic electron correlations, not captured by independent particle approaches, e.g. the
reduced single particle density matrix where we
discussed possible bias of previous calculations.
In order to describe the electronic structure of real materials, the periodic lattice potential due to the ionic crystal must be further included in the Hamiltonian. We expect \gls{wap}-net to still provide an accurate description to describe the electronic structure in materials, a detailed study to benchmark its performance for solids will be addressed in a future work.

After the ArXiv submission of our manuscript, two
related papers studying the electron gas with similar a neural network approach appeared, \cite{PhysRevLett.130.036401} focusing on the 
transition from the Fermi liquid to the Wigner crystal,
and  \cite{LiLi} towards more realistic systems. 

\section{Acknowledgements}

We thank Jan Hermann for useful early discussions and Hannah Auger for her expertise. The authors acknowledge support from the European Union's Horizon 2020 research and innovation program under Grant Agreement No. 957189 (BIG-MAP), No. 957213 (BATTERY2030PLUS), and No. 952165
(TREX). FW is thankful for support from NASA Academic Mission Services, Contract No. NNA16BD14C.
MH acknowledges the use of the Froggy platform of the
CIMENT infrastructure, which is supported by the Rh{\^o}ne-Alpes region (grant CPER07-13 CIRA)
 and the project Equip@Meso (ANR-10-EQPX-29-01) of the ANR.

\bibliography{bib}

\end{document}

%% file: glossary.tex
\newacronym{qmc}{QMC}{Quantum Monte Carlo}
\newacronym{mpc}{MPC}{something Periodic condition?}
\newacronym{mic}{MIC}{Minimum Image Convention}
\newacronym{kfac}{KFAC}{Kronecker Factored Approximate Curvature}
\newacronym{ib}{IB}{Iterative Backflow}
\newacronym{heg}{HEG}{Homogeneous Electron Gas}
\newacronym{hf}{HF}{Hartree Fock}
\newacronym{vmc}{VMC}{Variational Monte Carlo}
\newacronym{dmc}{DMC}{Diffusion Monte Carlo}
\newacronym{scf}{SCF}{Self-Consistent Field}
\newacronym{wap}{WAP}{Wave function Ansatz (but Periodic)}
\newacronym{gpu}{GPU}{Graphical processing units}

%% file: main.bbl
\begin{thebibliography}{77}%
\makeatletter
\providecommand \@ifxundefined [1]{%
 \@ifx{#1\undefined}
}%
\providecommand \@ifnum [1]{%
 \ifnum #1\expandafter \@firstoftwo
 \else \expandafter \@secondoftwo
 \fi
}%
\providecommand \@ifx [1]{%
 \ifx #1\expandafter \@firstoftwo
 \else \expandafter \@secondoftwo
 \fi
}%
\providecommand \natexlab [1]{#1}%
\providecommand \enquote  [1]{``#1''}%
\providecommand \bibnamefont  [1]{#1}%
\providecommand \bibfnamefont [1]{#1}%
\providecommand \citenamefont [1]{#1}%
\providecommand \href@noop [0]{\@secondoftwo}%
\providecommand \href [0]{\begingroup \@sanitize@url \@href}%
\providecommand \@href[1]{\@@startlink{#1}\@@href}%
\providecommand \@@href[1]{\endgroup#1\@@endlink}%
\providecommand \@sanitize@url [0]{\catcode `\\12\catcode `\$12\catcode
  `\&12\catcode `\#12\catcode `\^12\catcode `\_12\catcode `\%12\relax}%
\providecommand \@@startlink[1]{}%
\providecommand \@@endlink[0]{}%
\providecommand \url  [0]{\begingroup\@sanitize@url \@url }%
\providecommand \@url [1]{\endgroup\@href {#1}{\urlprefix }}%
\providecommand \urlprefix  [0]{URL }%
\providecommand \Eprint [0]{\href }%
\providecommand \doibase [0]{https://doi.org/}%
\providecommand \selectlanguage [0]{\@gobble}%
\providecommand \bibinfo  [0]{\@secondoftwo}%
\providecommand \bibfield  [0]{\@secondoftwo}%
\providecommand \translation [1]{[#1]}%
\providecommand \BibitemOpen [0]{}%
\providecommand \bibitemStop [0]{}%
\providecommand \bibitemNoStop [0]{.\EOS\space}%
\providecommand \EOS [0]{\spacefactor3000\relax}%
\providecommand \BibitemShut  [1]{\csname bibitem#1\endcsname}%
\let\auto@bib@innerbib\@empty
\bibitem [{\citenamefont {Martin}(2020)}]{martin2020electronic}%
  \BibitemOpen
  \bibfield  {author} {\bibinfo {author} {\bibfnamefont {R.~M.}\ \bibnamefont
  {Martin}},\ }\href@noop {} {\emph {\bibinfo {title} {Electronic structure:
  basic theory and practical methods}}}\ (\bibinfo  {publisher} {Cambridge
  university press},\ \bibinfo {year} {2020})\BibitemShut {NoStop}%
\bibitem [{\citenamefont {Pfau}\ \emph {et~al.}(2020)\citenamefont {Pfau},
  \citenamefont {Spencer}, \citenamefont {Matthews},\ and\ \citenamefont
  {Foulkes}}]{pfau2019ab}%
  \BibitemOpen
  \bibfield  {author} {\bibinfo {author} {\bibfnamefont {D.}~\bibnamefont
  {Pfau}}, \bibinfo {author} {\bibfnamefont {J.~S.}\ \bibnamefont {Spencer}},
  \bibinfo {author} {\bibfnamefont {A.~G. D.~G.}\ \bibnamefont {Matthews}},\
  and\ \bibinfo {author} {\bibfnamefont {W.~M.~C.}\ \bibnamefont {Foulkes}},\
  }\bibfield  {title} {\bibinfo {title} {Ab initio solution of the
  many-electron schr\"odinger equation with deep neural networks},\ }\href
  {https://doi.org/10.1103/PhysRevResearch.2.033429} {\bibfield  {journal}
  {\bibinfo  {journal} {Phys. Rev. Research}\ }\textbf {\bibinfo {volume}
  {2}},\ \bibinfo {pages} {033429} (\bibinfo {year} {2020})}\BibitemShut
  {NoStop}%
\bibitem [{\citenamefont {Hermann}\ \emph {et~al.}(2020)\citenamefont
  {Hermann}, \citenamefont {Sch{\"a}tzle},\ and\ \citenamefont
  {No{\'e}}}]{hermann2019deep}%
  \BibitemOpen
  \bibfield  {author} {\bibinfo {author} {\bibfnamefont {J.}~\bibnamefont
  {Hermann}}, \bibinfo {author} {\bibfnamefont {Z.}~\bibnamefont
  {Sch{\"a}tzle}},\ and\ \bibinfo {author} {\bibfnamefont {F.}~\bibnamefont
  {No{\'e}}},\ }\bibfield  {title} {\bibinfo {title} {Deep-neural-network
  solution of the electronic schr{\"o}dinger equation},\ }\href@noop {}
  {\bibfield  {journal} {\bibinfo  {journal} {Nature Chemistry}\ }\textbf
  {\bibinfo {volume} {12}},\ \bibinfo {pages} {891} (\bibinfo {year}
  {2020})}\BibitemShut {NoStop}%
\bibitem [{\citenamefont {Pescia}\ \emph {et~al.}(2021)\citenamefont {Pescia},
  \citenamefont {Han}, \citenamefont {Lovato}, \citenamefont {Lu},\ and\
  \citenamefont {Carleo}}]{pescia2021neural}%
  \BibitemOpen
  \bibfield  {author} {\bibinfo {author} {\bibfnamefont {G.}~\bibnamefont
  {Pescia}}, \bibinfo {author} {\bibfnamefont {J.}~\bibnamefont {Han}},
  \bibinfo {author} {\bibfnamefont {A.}~\bibnamefont {Lovato}}, \bibinfo
  {author} {\bibfnamefont {J.}~\bibnamefont {Lu}},\ and\ \bibinfo {author}
  {\bibfnamefont {G.}~\bibnamefont {Carleo}},\ }\bibfield  {title} {\bibinfo
  {title} {Neural-network quantum states for periodic systems in continuous
  space},\ }\href@noop {} {\bibfield  {journal} {\bibinfo  {journal} {arXiv
  preprint arXiv:2112.11957}\ } (\bibinfo {year} {2021})}\BibitemShut {NoStop}%
\bibitem [{\citenamefont {Holzmann}\ and\ \citenamefont {Moroni}(2019)}]{OBF}%
  \BibitemOpen
  \bibfield  {author} {\bibinfo {author} {\bibfnamefont {M.}~\bibnamefont
  {Holzmann}}\ and\ \bibinfo {author} {\bibfnamefont {S.}~\bibnamefont
  {Moroni}},\ }\bibfield  {title} {\bibinfo {title} {Orbital-dependent backflow
  wave functions for real-space quantum monte carlo},\ }\href
  {https://doi.org/10.1103/PhysRevB.99.085121} {\bibfield  {journal} {\bibinfo
  {journal} {Phys. Rev. B}\ }\textbf {\bibinfo {volume} {99}},\ \bibinfo
  {pages} {085121} (\bibinfo {year} {2019})}\BibitemShut {NoStop}%
\bibitem [{\citenamefont {Taddei}\ \emph {et~al.}(2015)\citenamefont {Taddei},
  \citenamefont {Ruggeri}, \citenamefont {Moroni},\ and\ \citenamefont
  {Holzmann}}]{taddei2015iterative}%
  \BibitemOpen
  \bibfield  {author} {\bibinfo {author} {\bibfnamefont {M.}~\bibnamefont
  {Taddei}}, \bibinfo {author} {\bibfnamefont {M.}~\bibnamefont {Ruggeri}},
  \bibinfo {author} {\bibfnamefont {S.}~\bibnamefont {Moroni}},\ and\ \bibinfo
  {author} {\bibfnamefont {M.}~\bibnamefont {Holzmann}},\ }\bibfield  {title}
  {\bibinfo {title} {Iterative backflow renormalization procedure for many-body
  ground-state wave functions of strongly interacting normal fermi liquids},\
  }\href@noop {} {\bibfield  {journal} {\bibinfo  {journal} {Physical Review
  B}\ }\textbf {\bibinfo {volume} {91}},\ \bibinfo {pages} {115106} (\bibinfo
  {year} {2015})}\BibitemShut {NoStop}%
\bibitem [{\citenamefont {Martin}\ \emph {et~al.}(2016)\citenamefont {Martin},
  \citenamefont {Reining},\ and\ \citenamefont
  {Ceperley}}]{martin2016interacting}%
  \BibitemOpen
  \bibfield  {author} {\bibinfo {author} {\bibfnamefont {R.~M.}\ \bibnamefont
  {Martin}}, \bibinfo {author} {\bibfnamefont {L.}~\bibnamefont {Reining}},\
  and\ \bibinfo {author} {\bibfnamefont {D.~M.}\ \bibnamefont {Ceperley}},\
  }\href@noop {} {\emph {\bibinfo {title} {Interacting electrons}}}\ (\bibinfo
  {publisher} {Cambridge University Press},\ \bibinfo {year}
  {2016})\BibitemShut {NoStop}%
\bibitem [{\citenamefont {Mahan}(2000)}]{mahan2000homogeneous}%
  \BibitemOpen
  \bibfield  {author} {\bibinfo {author} {\bibfnamefont {G.~D.}\ \bibnamefont
  {Mahan}},\ }\bibfield  {title} {\bibinfo {title} {Homogeneous electron gas},\
  }in\ \href@noop {} {\emph {\bibinfo {booktitle} {Many-Particle Physics}}}\
  (\bibinfo  {publisher} {Springer},\ \bibinfo {year} {2000})\ pp.\ \bibinfo
  {pages} {295--374}\BibitemShut {NoStop}%
\bibitem [{\citenamefont {Jacobsen}\ \emph {et~al.}(1987)\citenamefont
  {Jacobsen}, \citenamefont {Norskov},\ and\ \citenamefont
  {Puska}}]{jacobsen1987interatomic}%
  \BibitemOpen
  \bibfield  {author} {\bibinfo {author} {\bibfnamefont {K.~W.}\ \bibnamefont
  {Jacobsen}}, \bibinfo {author} {\bibfnamefont {J.}~\bibnamefont {Norskov}},\
  and\ \bibinfo {author} {\bibfnamefont {M.~J.}\ \bibnamefont {Puska}},\
  }\bibfield  {title} {\bibinfo {title} {Interatomic interactions in the
  effective-medium theory},\ }\href@noop {} {\bibfield  {journal} {\bibinfo
  {journal} {Physical Review B}\ }\textbf {\bibinfo {volume} {35}},\ \bibinfo
  {pages} {7423} (\bibinfo {year} {1987})}\BibitemShut {NoStop}%
\bibitem [{\citenamefont {Foulkes}\ \emph {et~al.}(2001)\citenamefont
  {Foulkes}, \citenamefont {Mitas}, \citenamefont {Needs},\ and\ \citenamefont
  {Rajagopal}}]{foulkes2001quantum}%
  \BibitemOpen
  \bibfield  {author} {\bibinfo {author} {\bibfnamefont {W.}~\bibnamefont
  {Foulkes}}, \bibinfo {author} {\bibfnamefont {L.}~\bibnamefont {Mitas}},
  \bibinfo {author} {\bibfnamefont {R.}~\bibnamefont {Needs}},\ and\ \bibinfo
  {author} {\bibfnamefont {G.}~\bibnamefont {Rajagopal}},\ }\bibfield  {title}
  {\bibinfo {title} {Quantum monte carlo simulations of solids},\ }\href@noop
  {} {\bibfield  {journal} {\bibinfo  {journal} {Reviews of Modern Physics}\
  }\textbf {\bibinfo {volume} {73}},\ \bibinfo {pages} {33} (\bibinfo {year}
  {2001})}\BibitemShut {NoStop}%
\bibitem [{\citenamefont {Harrison}(1980)}]{harrison1980solid}%
  \BibitemOpen
  \bibfield  {author} {\bibinfo {author} {\bibfnamefont {W.~A.}\ \bibnamefont
  {Harrison}},\ }\href@noop {} {\emph {\bibinfo {title} {Solid state theory}}}\
  (\bibinfo  {publisher} {Courier Corporation},\ \bibinfo {year}
  {1980})\BibitemShut {NoStop}%
\bibitem [{\citenamefont {Giuliani}\ and\ \citenamefont
  {Vignale}(2005)}]{giuliani2005quantum}%
  \BibitemOpen
  \bibfield  {author} {\bibinfo {author} {\bibfnamefont {G.}~\bibnamefont
  {Giuliani}}\ and\ \bibinfo {author} {\bibfnamefont {G.}~\bibnamefont
  {Vignale}},\ }\href@noop {} {\emph {\bibinfo {title} {Quantum theory of the
  electron liquid}}}\ (\bibinfo  {publisher} {Cambridge university press},\
  \bibinfo {year} {2005})\BibitemShut {NoStop}%
\bibitem [{\citenamefont {Hedin}(1965)}]{Hedin65}%
  \BibitemOpen
  \bibfield  {author} {\bibinfo {author} {\bibfnamefont {L.}~\bibnamefont
  {Hedin}},\ }\bibfield  {title} {\bibinfo {title} {New method for calculating
  the one-particle green's function with application to the electron-gas
  problem},\ }\href {https://doi.org/10.1103/PhysRev.139.A796} {\bibfield
  {journal} {\bibinfo  {journal} {Phys. Rev.}\ }\textbf {\bibinfo {volume}
  {139}},\ \bibinfo {pages} {A796} (\bibinfo {year} {1965})}\BibitemShut
  {NoStop}%
\bibitem [{\citenamefont {Chen}\ and\ \citenamefont
  {Haule}(2019)}]{chen2019combined}%
  \BibitemOpen
  \bibfield  {author} {\bibinfo {author} {\bibfnamefont {K.}~\bibnamefont
  {Chen}}\ and\ \bibinfo {author} {\bibfnamefont {K.}~\bibnamefont {Haule}},\
  }\bibfield  {title} {\bibinfo {title} {A combined variational and
  diagrammatic quantum monte carlo approach to the many-electron problem},\
  }\href@noop {} {\bibfield  {journal} {\bibinfo  {journal} {Nature
  communications}\ }\textbf {\bibinfo {volume} {10}},\ \bibinfo {pages} {1}
  (\bibinfo {year} {2019})}\BibitemShut {NoStop}%
\bibitem [{\citenamefont {Ceperley}(1978)}]{ceperley1978ground}%
  \BibitemOpen
  \bibfield  {author} {\bibinfo {author} {\bibfnamefont {D.}~\bibnamefont
  {Ceperley}},\ }\bibfield  {title} {\bibinfo {title} {Ground state of the
  fermion one-component plasma: A monte carlo study in two and three
  dimensions},\ }\href@noop {} {\bibfield  {journal} {\bibinfo  {journal}
  {Physical Review B}\ }\textbf {\bibinfo {volume} {18}},\ \bibinfo {pages}
  {3126} (\bibinfo {year} {1978})}\BibitemShut {NoStop}%
\bibitem [{\citenamefont {Ceperley}\ and\ \citenamefont
  {Alder}(1980)}]{ceperley1980ground}%
  \BibitemOpen
  \bibfield  {author} {\bibinfo {author} {\bibfnamefont {D.~M.}\ \bibnamefont
  {Ceperley}}\ and\ \bibinfo {author} {\bibfnamefont {B.~J.}\ \bibnamefont
  {Alder}},\ }\bibfield  {title} {\bibinfo {title} {Ground state of the
  electron gas by a stochastic method},\ }\href@noop {} {\bibfield  {journal}
  {\bibinfo  {journal} {Physical review letters}\ }\textbf {\bibinfo {volume}
  {45}},\ \bibinfo {pages} {566} (\bibinfo {year} {1980})}\BibitemShut
  {NoStop}%
\bibitem [{\citenamefont {Shepherd}\ \emph {et~al.}(2012)\citenamefont
  {Shepherd}, \citenamefont {Booth},\ and\ \citenamefont
  {Alavi}}]{shepherd2012investigation}%
  \BibitemOpen
  \bibfield  {author} {\bibinfo {author} {\bibfnamefont {J.~J.}\ \bibnamefont
  {Shepherd}}, \bibinfo {author} {\bibfnamefont {G.~H.}\ \bibnamefont
  {Booth}},\ and\ \bibinfo {author} {\bibfnamefont {A.}~\bibnamefont {Alavi}},\
  }\bibfield  {title} {\bibinfo {title} {Investigation of the full
  configuration interaction quantum monte carlo method using homogeneous
  electron gas models},\ }\href@noop {} {\bibfield  {journal} {\bibinfo
  {journal} {The Journal of chemical physics}\ }\textbf {\bibinfo {volume}
  {136}},\ \bibinfo {pages} {244101} (\bibinfo {year} {2012})}\BibitemShut
  {NoStop}%
\bibitem [{\citenamefont {Shepherd}\ and\ \citenamefont
  {Gr{\"u}neis}(2013)}]{shepherd2013many}%
  \BibitemOpen
  \bibfield  {author} {\bibinfo {author} {\bibfnamefont {J.~J.}\ \bibnamefont
  {Shepherd}}\ and\ \bibinfo {author} {\bibfnamefont {A.}~\bibnamefont
  {Gr{\"u}neis}},\ }\bibfield  {title} {\bibinfo {title} {Many-body quantum
  chemistry for the electron gas: Convergent perturbative theories},\
  }\href@noop {} {\bibfield  {journal} {\bibinfo  {journal} {Physical review
  letters}\ }\textbf {\bibinfo {volume} {110}},\ \bibinfo {pages} {226401}
  (\bibinfo {year} {2013})}\BibitemShut {NoStop}%
\bibitem [{\citenamefont {Ruggeri}\ \emph
  {et~al.}(2018{\natexlab{a}})\citenamefont {Ruggeri}, \citenamefont
  {R{\'\i}os},\ and\ \citenamefont {Alavi}}]{ruggeri2018correlation}%
  \BibitemOpen
  \bibfield  {author} {\bibinfo {author} {\bibfnamefont {M.}~\bibnamefont
  {Ruggeri}}, \bibinfo {author} {\bibfnamefont {P.~L.}\ \bibnamefont
  {R{\'\i}os}},\ and\ \bibinfo {author} {\bibfnamefont {A.}~\bibnamefont
  {Alavi}},\ }\bibfield  {title} {\bibinfo {title} {Correlation energies of the
  high-density spin-polarized electron gas to mev accuracy},\ }\href@noop {}
  {\bibfield  {journal} {\bibinfo  {journal} {Physical Review B}\ }\textbf
  {\bibinfo {volume} {98}},\ \bibinfo {pages} {161105} (\bibinfo {year}
  {2018}{\natexlab{a}})}\BibitemShut {NoStop}%
\bibitem [{\citenamefont {Raissi}\ \emph {et~al.}(2019)\citenamefont {Raissi},
  \citenamefont {Perdikaris},\ and\ \citenamefont
  {Karniadakis}}]{raissi2019physics}%
  \BibitemOpen
  \bibfield  {author} {\bibinfo {author} {\bibfnamefont {M.}~\bibnamefont
  {Raissi}}, \bibinfo {author} {\bibfnamefont {P.}~\bibnamefont {Perdikaris}},\
  and\ \bibinfo {author} {\bibfnamefont {G.~E.}\ \bibnamefont {Karniadakis}},\
  }\bibfield  {title} {\bibinfo {title} {Physics-informed neural networks: A
  deep learning framework for solving forward and inverse problems involving
  nonlinear partial differential equations},\ }\href@noop {} {\bibfield
  {journal} {\bibinfo  {journal} {Journal of Computational Physics}\ }\textbf
  {\bibinfo {volume} {378}},\ \bibinfo {pages} {686} (\bibinfo {year}
  {2019})}\BibitemShut {NoStop}%
\bibitem [{\citenamefont {Silver}\ \emph {et~al.}(2017)\citenamefont {Silver},
  \citenamefont {Schrittwieser}, \citenamefont {Simonyan}, \citenamefont
  {Antonoglou}, \citenamefont {Huang}, \citenamefont {Guez}, \citenamefont
  {Hubert}, \citenamefont {Baker}, \citenamefont {Lai}, \citenamefont {Bolton}
  \emph {et~al.}}]{silver2017mastering}%
  \BibitemOpen
  \bibfield  {author} {\bibinfo {author} {\bibfnamefont {D.}~\bibnamefont
  {Silver}}, \bibinfo {author} {\bibfnamefont {J.}~\bibnamefont
  {Schrittwieser}}, \bibinfo {author} {\bibfnamefont {K.}~\bibnamefont
  {Simonyan}}, \bibinfo {author} {\bibfnamefont {I.}~\bibnamefont
  {Antonoglou}}, \bibinfo {author} {\bibfnamefont {A.}~\bibnamefont {Huang}},
  \bibinfo {author} {\bibfnamefont {A.}~\bibnamefont {Guez}}, \bibinfo {author}
  {\bibfnamefont {T.}~\bibnamefont {Hubert}}, \bibinfo {author} {\bibfnamefont
  {L.}~\bibnamefont {Baker}}, \bibinfo {author} {\bibfnamefont
  {M.}~\bibnamefont {Lai}}, \bibinfo {author} {\bibfnamefont {A.}~\bibnamefont
  {Bolton}}, \emph {et~al.},\ }\bibfield  {title} {\bibinfo {title} {Mastering
  the game of go without human knowledge},\ }\href@noop {} {\bibfield
  {journal} {\bibinfo  {journal} {nature}\ }\textbf {\bibinfo {volume} {550}},\
  \bibinfo {pages} {354} (\bibinfo {year} {2017})}\BibitemShut {NoStop}%
\bibitem [{\citenamefont {Kiumarsi}\ \emph {et~al.}(2017)\citenamefont
  {Kiumarsi}, \citenamefont {Vamvoudakis}, \citenamefont {Modares},\ and\
  \citenamefont {Lewis}}]{kiumarsi2017optimal}%
  \BibitemOpen
  \bibfield  {author} {\bibinfo {author} {\bibfnamefont {B.}~\bibnamefont
  {Kiumarsi}}, \bibinfo {author} {\bibfnamefont {K.~G.}\ \bibnamefont
  {Vamvoudakis}}, \bibinfo {author} {\bibfnamefont {H.}~\bibnamefont
  {Modares}},\ and\ \bibinfo {author} {\bibfnamefont {F.~L.}\ \bibnamefont
  {Lewis}},\ }\bibfield  {title} {\bibinfo {title} {Optimal and autonomous
  control using reinforcement learning: A survey},\ }\href@noop {} {\bibfield
  {journal} {\bibinfo  {journal} {IEEE transactions on neural networks and
  learning systems}\ }\textbf {\bibinfo {volume} {29}},\ \bibinfo {pages}
  {2042} (\bibinfo {year} {2017})}\BibitemShut {NoStop}%
\bibitem [{\citenamefont {Carleo}\ and\ \citenamefont
  {Troyer}(2017)}]{carleoSolvingQuantumManyBody2017}%
  \BibitemOpen
  \bibfield  {author} {\bibinfo {author} {\bibfnamefont {G.}~\bibnamefont
  {Carleo}}\ and\ \bibinfo {author} {\bibfnamefont {M.}~\bibnamefont
  {Troyer}},\ }\bibfield  {title} {\bibinfo {title} {Solving the {{Quantum
  Many}}-{{Body Problem}} with {{Artificial Neural Networks}}},\ }\href
  {https://doi.org/10.1126/science.aag2302} {\bibfield  {journal} {\bibinfo
  {journal} {Science}\ }\textbf {\bibinfo {volume} {355}},\ \bibinfo {pages}
  {602} (\bibinfo {year} {2017})},\ \Eprint {https://arxiv.org/abs/1606.02318}
  {arXiv:1606.02318} \BibitemShut {NoStop}%
\bibitem [{\citenamefont {Wilson}\ \emph {et~al.}(2021)\citenamefont {Wilson},
  \citenamefont {Gao}, \citenamefont {Wudarski}, \citenamefont {Rieffel},\ and\
  \citenamefont {Tubman}}]{wilsonSimulationsStateoftheartFermionic2021}%
  \BibitemOpen
  \bibfield  {author} {\bibinfo {author} {\bibfnamefont {M.}~\bibnamefont
  {Wilson}}, \bibinfo {author} {\bibfnamefont {N.}~\bibnamefont {Gao}},
  \bibinfo {author} {\bibfnamefont {F.}~\bibnamefont {Wudarski}}, \bibinfo
  {author} {\bibfnamefont {E.}~\bibnamefont {Rieffel}},\ and\ \bibinfo {author}
  {\bibfnamefont {N.~M.}\ \bibnamefont {Tubman}},\ }\bibfield  {title}
  {\bibinfo {title} {Simulations of state-of-the-art fermionic neural network
  wave functions with diffusion {{Monte Carlo}}},\ }\href@noop {} {\bibfield
  {journal} {\bibinfo  {journal} {arXiv:2103.12570 [physics,
  physics:quant-ph]}\ } (\bibinfo {year} {2021})}\BibitemShut {NoStop}%
\bibitem [{\citenamefont {Gao}\ and\ \citenamefont
  {G{\"u}nnemann}(2022)}]{gao_pesnet_2022}%
  \BibitemOpen
  \bibfield  {author} {\bibinfo {author} {\bibfnamefont {N.}~\bibnamefont
  {Gao}}\ and\ \bibinfo {author} {\bibfnamefont {S.}~\bibnamefont
  {G{\"u}nnemann}},\ }\bibfield  {title} {\bibinfo {title} {Ab-initio potential
  energy surfaces by pairing gnns with neural wave functions},\ }in\ \href@noop
  {} {\emph {\bibinfo {booktitle} {International Conference on Learning
  Representations (ICLR)}}}\ (\bibinfo {year} {2022})\BibitemShut {NoStop}%
\bibitem [{\citenamefont {Spencer}\ \emph {et~al.}(2020)\citenamefont
  {Spencer}, \citenamefont {Pfau}, \citenamefont {Botev},\ and\ \citenamefont
  {Foulkes}}]{spencer2020better}%
  \BibitemOpen
  \bibfield  {author} {\bibinfo {author} {\bibfnamefont {J.~S.}\ \bibnamefont
  {Spencer}}, \bibinfo {author} {\bibfnamefont {D.}~\bibnamefont {Pfau}},
  \bibinfo {author} {\bibfnamefont {A.}~\bibnamefont {Botev}},\ and\ \bibinfo
  {author} {\bibfnamefont {W.~M.~C.}\ \bibnamefont {Foulkes}},\ }\bibfield
  {title} {\bibinfo {title} {Better, faster fermionic neural networks},\
  }\href@noop {} {\bibfield  {journal} {\bibinfo  {journal} {arXiv preprint
  arXiv:2011.07125}\ } (\bibinfo {year} {2020})}\BibitemShut {NoStop}%
\bibitem [{\citenamefont {Haule}\ and\ \citenamefont {Chen}(2022)}]{Dia}%
  \BibitemOpen
  \bibfield  {author} {\bibinfo {author} {\bibfnamefont {K.}~\bibnamefont
  {Haule}}\ and\ \bibinfo {author} {\bibfnamefont {K.}~\bibnamefont {Chen}},\
  }\bibfield  {title} {\bibinfo {title} {Single-particle excitations in the
  uniform electron gas by diagrammatic monte carlo},\ }\href
  {https://doi.org/10.1038/s41598-022-06188-6} {\bibfield  {journal} {\bibinfo
  {journal} {Scientific Reports}\ }\textbf {\bibinfo {volume} {12}},\ \bibinfo
  {pages} {2294} (\bibinfo {year} {2022})}\BibitemShut {NoStop}%
\bibitem [{\citenamefont {Holzmann}\ \emph {et~al.}(2011)\citenamefont
  {Holzmann}, \citenamefont {Bernu}, \citenamefont {Pierleoni}, \citenamefont
  {McMinis}, \citenamefont {Ceperley}, \citenamefont {Olevano},\ and\
  \citenamefont {Delle~Site}}]{momk3D}%
  \BibitemOpen
  \bibfield  {author} {\bibinfo {author} {\bibfnamefont {M.}~\bibnamefont
  {Holzmann}}, \bibinfo {author} {\bibfnamefont {B.}~\bibnamefont {Bernu}},
  \bibinfo {author} {\bibfnamefont {C.}~\bibnamefont {Pierleoni}}, \bibinfo
  {author} {\bibfnamefont {J.}~\bibnamefont {McMinis}}, \bibinfo {author}
  {\bibfnamefont {D.~M.}\ \bibnamefont {Ceperley}}, \bibinfo {author}
  {\bibfnamefont {V.}~\bibnamefont {Olevano}},\ and\ \bibinfo {author}
  {\bibfnamefont {L.}~\bibnamefont {Delle~Site}},\ }\bibfield  {title}
  {\bibinfo {title} {Momentum distribution of the homogeneous electron gas},\
  }\href {https://doi.org/10.1103/PhysRevLett.107.110402} {\bibfield  {journal}
  {\bibinfo  {journal} {Phys. Rev. Lett.}\ }\textbf {\bibinfo {volume} {107}},\
  \bibinfo {pages} {110402} (\bibinfo {year} {2011})}\BibitemShut {NoStop}%
\bibitem [{\citenamefont {von Glehn}\ \emph {et~al.}(2022)\citenamefont {von
  Glehn}, \citenamefont {Spencer},\ and\ \citenamefont {Pfau}}]{von2022self}%
  \BibitemOpen
  \bibfield  {author} {\bibinfo {author} {\bibfnamefont {I.}~\bibnamefont {von
  Glehn}}, \bibinfo {author} {\bibfnamefont {J.~S.}\ \bibnamefont {Spencer}},\
  and\ \bibinfo {author} {\bibfnamefont {D.}~\bibnamefont {Pfau}},\ }\bibfield
  {title} {\bibinfo {title} {A self-attention ansatz for ab-initio quantum
  chemistry},\ }\href@noop {} {\bibfield  {journal} {\bibinfo  {journal} {arXiv
  preprint arXiv:2211.13672}\ } (\bibinfo {year} {2022})}\BibitemShut {NoStop}%
\bibitem [{\citenamefont {Li}\ \emph {et~al.}(2022{\natexlab{a}})\citenamefont
  {Li}, \citenamefont {Li},\ and\ \citenamefont {Chen}}]{li2022ab}%
  \BibitemOpen
  \bibfield  {author} {\bibinfo {author} {\bibfnamefont {X.}~\bibnamefont
  {Li}}, \bibinfo {author} {\bibfnamefont {Z.}~\bibnamefont {Li}},\ and\
  \bibinfo {author} {\bibfnamefont {J.}~\bibnamefont {Chen}},\ }\bibfield
  {title} {\bibinfo {title} {Ab initio calculation of real solids via neural
  network ansatz},\ }\href@noop {} {\bibfield  {journal} {\bibinfo  {journal}
  {arXiv preprint arXiv:2203.15472}\ } (\bibinfo {year}
  {2022}{\natexlab{a}})}\BibitemShut {NoStop}%
\bibitem [{\citenamefont {Schmidt}\ \emph {et~al.}(1981)\citenamefont
  {Schmidt}, \citenamefont {Lee}, \citenamefont {Kalos},\ and\ \citenamefont
  {Chester}}]{Schmidt81}%
  \BibitemOpen
  \bibfield  {author} {\bibinfo {author} {\bibfnamefont {K.~E.}\ \bibnamefont
  {Schmidt}}, \bibinfo {author} {\bibfnamefont {M.~A.}\ \bibnamefont {Lee}},
  \bibinfo {author} {\bibfnamefont {M.~H.}\ \bibnamefont {Kalos}},\ and\
  \bibinfo {author} {\bibfnamefont {G.~V.}\ \bibnamefont {Chester}},\
  }\bibfield  {title} {\bibinfo {title} {Structure of the ground state of a
  fermion fluid},\ }\href {https://doi.org/10.1103/PhysRevLett.47.807}
  {\bibfield  {journal} {\bibinfo  {journal} {Phys. Rev. Lett.}\ }\textbf
  {\bibinfo {volume} {47}},\ \bibinfo {pages} {807} (\bibinfo {year}
  {1981})}\BibitemShut {NoStop}%
\bibitem [{\citenamefont {Holzmann}\ \emph {et~al.}(2006)\citenamefont
  {Holzmann}, \citenamefont {Bernu},\ and\ \citenamefont {Ceperley}}]{BFN}%
  \BibitemOpen
  \bibfield  {author} {\bibinfo {author} {\bibfnamefont {M.}~\bibnamefont
  {Holzmann}}, \bibinfo {author} {\bibfnamefont {B.}~\bibnamefont {Bernu}},\
  and\ \bibinfo {author} {\bibfnamefont {D.~M.}\ \bibnamefont {Ceperley}},\
  }\bibfield  {title} {\bibinfo {title} {Many-body wavefunctions for normal
  liquid $^{3}\mathrm{He}$},\ }\href
  {https://doi.org/10.1103/PhysRevB.74.104510} {\bibfield  {journal} {\bibinfo
  {journal} {Phys. Rev. B}\ }\textbf {\bibinfo {volume} {74}},\ \bibinfo
  {pages} {104510} (\bibinfo {year} {2006})}\BibitemShut {NoStop}%
\bibitem [{\citenamefont {Bouab{\c{c}}a}\ \emph {et~al.}(2010)\citenamefont
  {Bouab{\c{c}}a}, \citenamefont {Bra{\"\i}da},\ and\ \citenamefont
  {Caffarel}}]{caffarel10}%
  \BibitemOpen
  \bibfield  {author} {\bibinfo {author} {\bibfnamefont {T.}~\bibnamefont
  {Bouab{\c{c}}a}}, \bibinfo {author} {\bibfnamefont {B.}~\bibnamefont
  {Bra{\"\i}da}},\ and\ \bibinfo {author} {\bibfnamefont {M.}~\bibnamefont
  {Caffarel}},\ }\bibfield  {title} {\bibinfo {title} {Multi-jastrow trial
  wavefunctions for electronic structure calculations with quantum monte
  carlo},\ }\href@noop {} {\bibfield  {journal} {\bibinfo  {journal} {The
  Journal of chemical physics}\ }\textbf {\bibinfo {volume} {133}},\ \bibinfo
  {pages} {044111} (\bibinfo {year} {2010})}\BibitemShut {NoStop}%
\bibitem [{\citenamefont {Ruggeri}\ \emph
  {et~al.}(2018{\natexlab{b}})\citenamefont {Ruggeri}, \citenamefont {Moroni},\
  and\ \citenamefont {Holzmann}}]{ruggeri2018nonlinear}%
  \BibitemOpen
  \bibfield  {author} {\bibinfo {author} {\bibfnamefont {M.}~\bibnamefont
  {Ruggeri}}, \bibinfo {author} {\bibfnamefont {S.}~\bibnamefont {Moroni}},\
  and\ \bibinfo {author} {\bibfnamefont {M.}~\bibnamefont {Holzmann}},\
  }\bibfield  {title} {\bibinfo {title} {Nonlinear network description for
  many-body quantum systems in continuous space},\ }\href@noop {} {\bibfield
  {journal} {\bibinfo  {journal} {Physical review letters}\ }\textbf {\bibinfo
  {volume} {120}},\ \bibinfo {pages} {205302} (\bibinfo {year}
  {2018}{\natexlab{b}})}\BibitemShut {NoStop}%
\bibitem [{\citenamefont {Holzmann}\ and\ \citenamefont
  {Moroni}(2020)}]{BFStoner}%
  \BibitemOpen
  \bibfield  {author} {\bibinfo {author} {\bibfnamefont {M.}~\bibnamefont
  {Holzmann}}\ and\ \bibinfo {author} {\bibfnamefont {S.}~\bibnamefont
  {Moroni}},\ }\bibfield  {title} {\bibinfo {title} {Itinerant-electron
  magnetism: The importance of many-body correlations},\ }\href
  {https://doi.org/10.1103/PhysRevLett.124.206404} {\bibfield  {journal}
  {\bibinfo  {journal} {Phys. Rev. Lett.}\ }\textbf {\bibinfo {volume} {124}},\
  \bibinfo {pages} {206404} (\bibinfo {year} {2020})}\BibitemShut {NoStop}%
\bibitem [{\citenamefont {Kessler}\ \emph {et~al.}(2021)\citenamefont
  {Kessler}, \citenamefont {Calcavecchia},\ and\ \citenamefont
  {K{\"u}hne}}]{kessler2021artificial}%
  \BibitemOpen
  \bibfield  {author} {\bibinfo {author} {\bibfnamefont {J.}~\bibnamefont
  {Kessler}}, \bibinfo {author} {\bibfnamefont {F.}~\bibnamefont
  {Calcavecchia}},\ and\ \bibinfo {author} {\bibfnamefont {T.~D.}\ \bibnamefont
  {K{\"u}hne}},\ }\bibfield  {title} {\bibinfo {title} {Artificial neural
  networks as trial wave functions for quantum monte carlo},\ }\href@noop {}
  {\bibfield  {journal} {\bibinfo  {journal} {Advanced Theory and Simulations}\
  }\textbf {\bibinfo {volume} {4}},\ \bibinfo {pages} {2000269} (\bibinfo
  {year} {2021})}\BibitemShut {NoStop}%
\bibitem [{\citenamefont {Han}\ \emph {et~al.}(2019)\citenamefont {Han},
  \citenamefont {Zhang},\ and\ \citenamefont
  {E}}]{hanSolvingManyelectronSchrodinger2019}%
  \BibitemOpen
  \bibfield  {author} {\bibinfo {author} {\bibfnamefont {J.}~\bibnamefont
  {Han}}, \bibinfo {author} {\bibfnamefont {L.}~\bibnamefont {Zhang}},\ and\
  \bibinfo {author} {\bibfnamefont {W.}~\bibnamefont {E}},\ }\bibfield  {title}
  {\bibinfo {title} {Solving many-electron {{Schr\"odinger}} equation using
  deep neural networks},\ }\href {https://doi.org/10.1016/j.jcp.2019.108929}
  {\bibfield  {journal} {\bibinfo  {journal} {Journal of Computational
  Physics}\ }\textbf {\bibinfo {volume} {399}},\ \bibinfo {pages} {108929}
  (\bibinfo {year} {2019})}\BibitemShut {NoStop}%
\bibitem [{\citenamefont {Choo}\ \emph {et~al.}(2020)\citenamefont {Choo},
  \citenamefont {Mezzacapo},\ and\ \citenamefont
  {Carleo}}]{chooFermionicNeuralnetworkStates2020}%
  \BibitemOpen
  \bibfield  {author} {\bibinfo {author} {\bibfnamefont {K.}~\bibnamefont
  {Choo}}, \bibinfo {author} {\bibfnamefont {A.}~\bibnamefont {Mezzacapo}},\
  and\ \bibinfo {author} {\bibfnamefont {G.}~\bibnamefont {Carleo}},\
  }\bibfield  {title} {\bibinfo {title} {Fermionic neural-network states for
  ab-initio electronic structure},\ }\href
  {https://doi.org/10.1038/s41467-020-15724-9} {\bibfield  {journal} {\bibinfo
  {journal} {Nature Communications}\ }\textbf {\bibinfo {volume} {11}},\
  \bibinfo {pages} {2368} (\bibinfo {year} {2020})}\BibitemShut {NoStop}%
\bibitem [{\citenamefont {Acevedo}\ \emph {et~al.}(2020)\citenamefont
  {Acevedo}, \citenamefont {Curry}, \citenamefont {Joshi}, \citenamefont
  {Leroux},\ and\ \citenamefont {Malaya}}]{acevedoVandermondeWaveFunction2020}%
  \BibitemOpen
  \bibfield  {author} {\bibinfo {author} {\bibfnamefont {A.}~\bibnamefont
  {Acevedo}}, \bibinfo {author} {\bibfnamefont {M.}~\bibnamefont {Curry}},
  \bibinfo {author} {\bibfnamefont {S.~H.}\ \bibnamefont {Joshi}}, \bibinfo
  {author} {\bibfnamefont {B.}~\bibnamefont {Leroux}},\ and\ \bibinfo {author}
  {\bibfnamefont {N.}~\bibnamefont {Malaya}},\ }\bibfield  {title} {\bibinfo
  {title} {Vandermonde {{Wave Function Ansatz}} for {{Improved Variational
  Monte Carlo}}},\ }in\ \href {https://doi.org/10.1109/DLS51937.2020.00010}
  {\emph {\bibinfo {booktitle} {2020 {{IEEE}}/{{ACM Fourth Workshop}} on {{Deep
  Learning}} on {{Supercomputers}} ({{DLS}})}}}\ (\bibinfo {year} {2020})\ pp.\
  \bibinfo {pages} {40--47}\BibitemShut {NoStop}%
\bibitem [{\citenamefont {Li}\ \emph {et~al.}(2021)\citenamefont {Li},
  \citenamefont {Zhai},\ and\ \citenamefont
  {Chen}}]{liNeuralnetworkbasedMultistateSolver2021}%
  \BibitemOpen
  \bibfield  {author} {\bibinfo {author} {\bibfnamefont {H.}~\bibnamefont
  {Li}}, \bibinfo {author} {\bibfnamefont {Q.}~\bibnamefont {Zhai}},\ and\
  \bibinfo {author} {\bibfnamefont {J.~Z.~Y.}\ \bibnamefont {Chen}},\
  }\bibfield  {title} {\bibinfo {title} {Neural-network-based multistate solver
  for a static {{Schr\"odinger}} equation},\ }\href
  {https://doi.org/10.1103/PhysRevA.103.032405} {\bibfield  {journal} {\bibinfo
   {journal} {Physical Review A}\ }\textbf {\bibinfo {volume} {103}},\ \bibinfo
  {pages} {032405} (\bibinfo {year} {2021})}\BibitemShut {NoStop}%
\bibitem [{\citenamefont {Wang}\ \emph {et~al.}(2022)\citenamefont {Wang},
  \citenamefont {Xie},\ and\ \citenamefont {Zhang}}]{wang2022m}%
  \BibitemOpen
  \bibfield  {author} {\bibinfo {author} {\bibfnamefont {L.}~\bibnamefont
  {Wang}}, \bibinfo {author} {\bibfnamefont {H.}~\bibnamefont {Xie}},\ and\
  \bibinfo {author} {\bibfnamefont {L.}~\bibnamefont {Zhang}},\ }\bibfield
  {title} {\bibinfo {title} {m* of electron gases: a neural canonical
  transformation study},\ }\href@noop {} {\bibfield  {journal} {\bibinfo
  {journal} {Bulletin of the American Physical Society}\ } (\bibinfo {year}
  {2022})}\BibitemShut {NoStop}%
\bibitem [{\citenamefont {Ziyin}\ \emph {et~al.}(2020)\citenamefont {Ziyin},
  \citenamefont {Hartwig},\ and\ \citenamefont {Ueda}}]{ziyin2020neural}%
  \BibitemOpen
  \bibfield  {author} {\bibinfo {author} {\bibfnamefont {L.}~\bibnamefont
  {Ziyin}}, \bibinfo {author} {\bibfnamefont {T.}~\bibnamefont {Hartwig}},\
  and\ \bibinfo {author} {\bibfnamefont {M.}~\bibnamefont {Ueda}},\ }\bibfield
  {title} {\bibinfo {title} {Neural networks fail to learn periodic functions
  and how to fix it},\ }\href@noop {} {\bibfield  {journal} {\bibinfo
  {journal} {arXiv preprint arXiv:2006.08195}\ } (\bibinfo {year}
  {2020})}\BibitemShut {NoStop}%
\bibitem [{\citenamefont {Sitzmann}\ \emph {et~al.}(2020)\citenamefont
  {Sitzmann}, \citenamefont {Martel}, \citenamefont {Bergman}, \citenamefont
  {Lindell},\ and\ \citenamefont {Wetzstein}}]{sitzmann2020implicit}%
  \BibitemOpen
  \bibfield  {author} {\bibinfo {author} {\bibfnamefont {V.}~\bibnamefont
  {Sitzmann}}, \bibinfo {author} {\bibfnamefont {J.}~\bibnamefont {Martel}},
  \bibinfo {author} {\bibfnamefont {A.}~\bibnamefont {Bergman}}, \bibinfo
  {author} {\bibfnamefont {D.}~\bibnamefont {Lindell}},\ and\ \bibinfo {author}
  {\bibfnamefont {G.}~\bibnamefont {Wetzstein}},\ }\bibfield  {title} {\bibinfo
  {title} {Implicit neural representations with periodic activation
  functions},\ }\href@noop {} {\bibfield  {journal} {\bibinfo  {journal}
  {Advances in Neural Information Processing Systems}\ }\textbf {\bibinfo
  {volume} {33}} (\bibinfo {year} {2020})}\BibitemShut {NoStop}%
\bibitem [{\citenamefont {Pescia}\ \emph {et~al.}(2022)\citenamefont {Pescia},
  \citenamefont {Han}, \citenamefont {Lovato}, \citenamefont {Lu},\ and\
  \citenamefont {Carleo}}]{Pescia}%
  \BibitemOpen
  \bibfield  {author} {\bibinfo {author} {\bibfnamefont {G.}~\bibnamefont
  {Pescia}}, \bibinfo {author} {\bibfnamefont {J.}~\bibnamefont {Han}},
  \bibinfo {author} {\bibfnamefont {A.}~\bibnamefont {Lovato}}, \bibinfo
  {author} {\bibfnamefont {J.}~\bibnamefont {Lu}},\ and\ \bibinfo {author}
  {\bibfnamefont {G.}~\bibnamefont {Carleo}},\ }\bibfield  {title} {\bibinfo
  {title} {Neural-network quantum states for periodic systems in continuous
  space},\ }\href {https://doi.org/10.1103/PhysRevResearch.4.023138} {\bibfield
   {journal} {\bibinfo  {journal} {Phys. Rev. Res.}\ }\textbf {\bibinfo
  {volume} {4}},\ \bibinfo {pages} {023138} (\bibinfo {year}
  {2022})}\BibitemShut {NoStop}%
\bibitem [{\citenamefont {Sch{\"u}tt}\ \emph {et~al.}(2018)\citenamefont
  {Sch{\"u}tt}, \citenamefont {Sauceda}, \citenamefont {Kindermans},
  \citenamefont {Tkatchenko},\ and\ \citenamefont
  {M{\"u}ller}}]{schutt2018schnet}%
  \BibitemOpen
  \bibfield  {author} {\bibinfo {author} {\bibfnamefont {K.~T.}\ \bibnamefont
  {Sch{\"u}tt}}, \bibinfo {author} {\bibfnamefont {H.~E.}\ \bibnamefont
  {Sauceda}}, \bibinfo {author} {\bibfnamefont {P.-J.}\ \bibnamefont
  {Kindermans}}, \bibinfo {author} {\bibfnamefont {A.}~\bibnamefont
  {Tkatchenko}},\ and\ \bibinfo {author} {\bibfnamefont {K.-R.}\ \bibnamefont
  {M{\"u}ller}},\ }\bibfield  {title} {\bibinfo {title} {Schnet--a deep
  learning architecture for molecules and materials},\ }\href@noop {}
  {\bibfield  {journal} {\bibinfo  {journal} {The Journal of Chemical Physics}\
  }\textbf {\bibinfo {volume} {148}},\ \bibinfo {pages} {241722} (\bibinfo
  {year} {2018})}\BibitemShut {NoStop}%
\bibitem [{\citenamefont {Xie}\ \emph {et~al.}(2021)\citenamefont {Xie},
  \citenamefont {Fu}, \citenamefont {Ganea}, \citenamefont {Barzilay},\ and\
  \citenamefont {Jaakkola}}]{xie2021crystal}%
  \BibitemOpen
  \bibfield  {author} {\bibinfo {author} {\bibfnamefont {T.}~\bibnamefont
  {Xie}}, \bibinfo {author} {\bibfnamefont {X.}~\bibnamefont {Fu}}, \bibinfo
  {author} {\bibfnamefont {O.-E.}\ \bibnamefont {Ganea}}, \bibinfo {author}
  {\bibfnamefont {R.}~\bibnamefont {Barzilay}},\ and\ \bibinfo {author}
  {\bibfnamefont {T.}~\bibnamefont {Jaakkola}},\ }\bibfield  {title} {\bibinfo
  {title} {Crystal diffusion variational autoencoder for periodic material
  generation},\ }\href@noop {} {\bibfield  {journal} {\bibinfo  {journal}
  {arXiv preprint arXiv:2110.06197}\ } (\bibinfo {year} {2021})}\BibitemShut
  {NoStop}%
\bibitem [{\citenamefont {Hornik}\ \emph {et~al.}(1989)\citenamefont {Hornik},
  \citenamefont {Stinchcombe},\ and\ \citenamefont
  {White}}]{hornik1989multilayer}%
  \BibitemOpen
  \bibfield  {author} {\bibinfo {author} {\bibfnamefont {K.}~\bibnamefont
  {Hornik}}, \bibinfo {author} {\bibfnamefont {M.}~\bibnamefont
  {Stinchcombe}},\ and\ \bibinfo {author} {\bibfnamefont {H.}~\bibnamefont
  {White}},\ }\bibfield  {title} {\bibinfo {title} {Multilayer feedforward
  networks are universal approximators},\ }\href@noop {} {\bibfield  {journal}
  {\bibinfo  {journal} {Neural networks}\ }\textbf {\bibinfo {volume} {2}},\
  \bibinfo {pages} {359} (\bibinfo {year} {1989})}\BibitemShut {NoStop}%
\bibitem [{\citenamefont {Ewald}(1921)}]{Ewald}%
  \BibitemOpen
  \bibfield  {author} {\bibinfo {author} {\bibfnamefont {P.~P.}\ \bibnamefont
  {Ewald}},\ }\bibfield  {title} {\bibinfo {title} {Die berechnung optischer
  und elektrostatischer gitterpotentiale},\ }\href
  {https://doi.org/https://doi.org/10.1002/andp.19213690304} {\bibfield
  {journal} {\bibinfo  {journal} {Annalen der Physik}\ }\textbf {\bibinfo
  {volume} {369}},\ \bibinfo {pages} {253} (\bibinfo {year}
  {1921})}\BibitemShut {NoStop}%
\bibitem [{\citenamefont {Szabo}\ and\ \citenamefont
  {Ostlund}(2012)}]{szabo2012modern}%
  \BibitemOpen
  \bibfield  {author} {\bibinfo {author} {\bibfnamefont {A.}~\bibnamefont
  {Szabo}}\ and\ \bibinfo {author} {\bibfnamefont {N.~S.}\ \bibnamefont
  {Ostlund}},\ }\href@noop {} {\emph {\bibinfo {title} {Modern quantum
  chemistry: introduction to advanced electronic structure theory}}}\ (\bibinfo
   {publisher} {Courier Corporation},\ \bibinfo {year} {2012})\BibitemShut
  {NoStop}%
\bibitem [{\citenamefont {Hastings}(1970)}]{hastings1970monte}%
  \BibitemOpen
  \bibfield  {author} {\bibinfo {author} {\bibfnamefont {W.~K.}\ \bibnamefont
  {Hastings}},\ }\bibfield  {title} {\bibinfo {title} {Monte carlo sampling
  methods using markov chains and their applications},\ }\href@noop {}
  {\bibfield  {journal} {\bibinfo  {journal} {Oxford University Press}\ }
  (\bibinfo {year} {1970})}\BibitemShut {NoStop}%
\bibitem [{\citenamefont {Suzuki}(1993)}]{suzuki1993quantum}%
  \BibitemOpen
  \bibfield  {author} {\bibinfo {author} {\bibfnamefont {M.}~\bibnamefont
  {Suzuki}},\ }\href@noop {} {\emph {\bibinfo {title} {Quantum Monte Carlo
  methods in condensed matter physics}}}\ (\bibinfo  {publisher} {World
  scientific},\ \bibinfo {year} {1993})\BibitemShut {NoStop}%
\bibitem [{\citenamefont {Lester~Jr}\ \emph {et~al.}(2009)\citenamefont
  {Lester~Jr}, \citenamefont {Mitas},\ and\ \citenamefont
  {Hammond}}]{lester2009quantum}%
  \BibitemOpen
  \bibfield  {author} {\bibinfo {author} {\bibfnamefont {W.~A.}\ \bibnamefont
  {Lester~Jr}}, \bibinfo {author} {\bibfnamefont {L.}~\bibnamefont {Mitas}},\
  and\ \bibinfo {author} {\bibfnamefont {B.}~\bibnamefont {Hammond}},\
  }\bibfield  {title} {\bibinfo {title} {Quantum monte carlo for atoms,
  molecules and solids},\ }\href@noop {} {\bibfield  {journal} {\bibinfo
  {journal} {Chemical Physics Letters}\ }\textbf {\bibinfo {volume} {478}},\
  \bibinfo {pages} {1} (\bibinfo {year} {2009})}\BibitemShut {NoStop}%
\bibitem [{\citenamefont {Elfwing}\ \emph {et~al.}(2018)\citenamefont
  {Elfwing}, \citenamefont {Uchibe},\ and\ \citenamefont
  {Doya}}]{elfwing2018sigmoid}%
  \BibitemOpen
  \bibfield  {author} {\bibinfo {author} {\bibfnamefont {S.}~\bibnamefont
  {Elfwing}}, \bibinfo {author} {\bibfnamefont {E.}~\bibnamefont {Uchibe}},\
  and\ \bibinfo {author} {\bibfnamefont {K.}~\bibnamefont {Doya}},\ }\bibfield
  {title} {\bibinfo {title} {Sigmoid-weighted linear units for neural network
  function approximation in reinforcement learning},\ }\href@noop {} {\bibfield
   {journal} {\bibinfo  {journal} {Neural Networks}\ }\textbf {\bibinfo
  {volume} {107}},\ \bibinfo {pages} {3} (\bibinfo {year} {2018})}\BibitemShut
  {NoStop}%
\bibitem [{\citenamefont {Bradbury}\ \emph {et~al.}(2018)\citenamefont
  {Bradbury}, \citenamefont {Frostig}, \citenamefont {Hawkins}, \citenamefont
  {Johnson}, \citenamefont {Leary}, \citenamefont {Maclaurin}, \citenamefont
  {Necula}, \citenamefont {Paszke}, \citenamefont {Vander{P}las}, \citenamefont
  {Wanderman-{M}ilne},\ and\ \citenamefont {Zhang}}]{jax2018github}%
  \BibitemOpen
  \bibfield  {author} {\bibinfo {author} {\bibfnamefont {J.}~\bibnamefont
  {Bradbury}}, \bibinfo {author} {\bibfnamefont {R.}~\bibnamefont {Frostig}},
  \bibinfo {author} {\bibfnamefont {P.}~\bibnamefont {Hawkins}}, \bibinfo
  {author} {\bibfnamefont {M.~J.}\ \bibnamefont {Johnson}}, \bibinfo {author}
  {\bibfnamefont {C.}~\bibnamefont {Leary}}, \bibinfo {author} {\bibfnamefont
  {D.}~\bibnamefont {Maclaurin}}, \bibinfo {author} {\bibfnamefont
  {G.}~\bibnamefont {Necula}}, \bibinfo {author} {\bibfnamefont
  {A.}~\bibnamefont {Paszke}}, \bibinfo {author} {\bibfnamefont
  {J.}~\bibnamefont {Vander{P}las}}, \bibinfo {author} {\bibfnamefont
  {S.}~\bibnamefont {Wanderman-{M}ilne}},\ and\ \bibinfo {author}
  {\bibfnamefont {Q.}~\bibnamefont {Zhang}},\ }\href
  {http://github.com/google/jax} {\bibinfo {title} {{JAX}: composable
  transformations of {P}ython+{N}um{P}y programs}} (\bibinfo {year}
  {2018})\BibitemShut {NoStop}%
\bibitem [{\citenamefont {Grosse}\ and\ \citenamefont
  {Martens}(2016)}]{grosse2016kronecker}%
  \BibitemOpen
  \bibfield  {author} {\bibinfo {author} {\bibfnamefont {R.}~\bibnamefont
  {Grosse}}\ and\ \bibinfo {author} {\bibfnamefont {J.}~\bibnamefont
  {Martens}},\ }\bibfield  {title} {\bibinfo {title} {A kronecker-factored
  approximate fisher matrix for convolution layers},\ }in\ \href@noop {} {\emph
  {\bibinfo {booktitle} {International Conference on Machine Learning}}}\
  (\bibinfo {organization} {PMLR},\ \bibinfo {year} {2016})\ pp.\ \bibinfo
  {pages} {573--582}\BibitemShut {NoStop}%
\bibitem [{\citenamefont {Martens}\ and\ \citenamefont
  {Grosse}(2015)}]{martens2015optimizing}%
  \BibitemOpen
  \bibfield  {author} {\bibinfo {author} {\bibfnamefont {J.}~\bibnamefont
  {Martens}}\ and\ \bibinfo {author} {\bibfnamefont {R.}~\bibnamefont
  {Grosse}},\ }\bibfield  {title} {\bibinfo {title} {Optimizing neural networks
  with kronecker-factored approximate curvature},\ }in\ \href@noop {} {\emph
  {\bibinfo {booktitle} {International conference on machine learning}}}\
  (\bibinfo {organization} {PMLR},\ \bibinfo {year} {2015})\ pp.\ \bibinfo
  {pages} {2408--2417}\BibitemShut {NoStop}%
\bibitem [{\citenamefont {Ba}\ \emph {et~al.}(2016)\citenamefont {Ba},
  \citenamefont {Grosse},\ and\ \citenamefont {Martens}}]{ba2016distributed}%
  \BibitemOpen
  \bibfield  {author} {\bibinfo {author} {\bibfnamefont {J.}~\bibnamefont
  {Ba}}, \bibinfo {author} {\bibfnamefont {R.}~\bibnamefont {Grosse}},\ and\
  \bibinfo {author} {\bibfnamefont {J.}~\bibnamefont {Martens}},\ }\bibfield
  {title} {\bibinfo {title} {Distributed second-order optimization using
  kronecker-factored approximations},\ }\href@noop {} {\  (\bibinfo {year}
  {2016})}\BibitemShut {NoStop}%
\bibitem [{\citenamefont {Sorella}(2001)}]{Sandro2001}%
  \BibitemOpen
  \bibfield  {author} {\bibinfo {author} {\bibfnamefont {S.}~\bibnamefont
  {Sorella}},\ }\bibfield  {title} {\bibinfo {title} {Generalized lanczos
  algorithm for variational quantum monte carlo},\ }\href
  {https://doi.org/10.1103/PhysRevB.64.024512} {\bibfield  {journal} {\bibinfo
  {journal} {Phys. Rev. B}\ }\textbf {\bibinfo {volume} {64}},\ \bibinfo
  {pages} {024512} (\bibinfo {year} {2001})}\BibitemShut {NoStop}%
\bibitem [{\citenamefont {Liao}\ \emph {et~al.}(2021)\citenamefont {Liao},
  \citenamefont {Schraivogel}, \citenamefont {Luo}, \citenamefont {Kats},\ and\
  \citenamefont {Alavi}}]{Transcorr}%
  \BibitemOpen
  \bibfield  {author} {\bibinfo {author} {\bibfnamefont {K.}~\bibnamefont
  {Liao}}, \bibinfo {author} {\bibfnamefont {T.}~\bibnamefont {Schraivogel}},
  \bibinfo {author} {\bibfnamefont {H.}~\bibnamefont {Luo}}, \bibinfo {author}
  {\bibfnamefont {D.}~\bibnamefont {Kats}},\ and\ \bibinfo {author}
  {\bibfnamefont {A.}~\bibnamefont {Alavi}},\ }\bibfield  {title} {\bibinfo
  {title} {Towards efficient and accurate ab initio solutions to periodic
  systems via transcorrelation and coupled cluster theory},\ }\href
  {https://doi.org/10.1103/PhysRevResearch.3.033072} {\bibfield  {journal}
  {\bibinfo  {journal} {Phys. Rev. Research}\ }\textbf {\bibinfo {volume}
  {3}},\ \bibinfo {pages} {033072} (\bibinfo {year} {2021})}\BibitemShut
  {NoStop}%
\bibitem [{\citenamefont {Brown}\ \emph {et~al.}(2007)\citenamefont {Brown},
  \citenamefont {Trail}, \citenamefont {López~Ríos},\ and\ \citenamefont
  {Needs}}]{brown07}%
  \BibitemOpen
  \bibfield  {author} {\bibinfo {author} {\bibfnamefont {M.~D.}\ \bibnamefont
  {Brown}}, \bibinfo {author} {\bibfnamefont {J.~R.}\ \bibnamefont {Trail}},
  \bibinfo {author} {\bibfnamefont {P.}~\bibnamefont {López~Ríos}},\ and\
  \bibinfo {author} {\bibfnamefont {R.~J.}\ \bibnamefont {Needs}},\ }\bibfield
  {title} {\bibinfo {title} {Energies of the first row atoms from quantum monte
  carlo},\ }\href {https://doi.org/10.1063/1.2743972} {\bibfield  {journal}
  {\bibinfo  {journal} {The Journal of Chemical Physics}\ }\textbf {\bibinfo
  {volume} {126}},\ \bibinfo {pages} {224110} (\bibinfo {year} {2007})},\
  \Eprint {https://arxiv.org/abs/https://doi.org/10.1063/1.2743972}
  {https://doi.org/10.1063/1.2743972} \BibitemShut {NoStop}%
\bibitem [{\citenamefont {Scherbela}\ \emph {et~al.}(2021)\citenamefont
  {Scherbela}, \citenamefont {Reisenhofer}, \citenamefont {Gerard},
  \citenamefont {Marquetand},\ and\ \citenamefont
  {Grohs}}]{scherbelaSolvingElectronicSchrodinger2021}%
  \BibitemOpen
  \bibfield  {author} {\bibinfo {author} {\bibfnamefont {M.}~\bibnamefont
  {Scherbela}}, \bibinfo {author} {\bibfnamefont {R.}~\bibnamefont
  {Reisenhofer}}, \bibinfo {author} {\bibfnamefont {L.}~\bibnamefont {Gerard}},
  \bibinfo {author} {\bibfnamefont {P.}~\bibnamefont {Marquetand}},\ and\
  \bibinfo {author} {\bibfnamefont {P.}~\bibnamefont {Grohs}},\ }\bibfield
  {title} {\bibinfo {title} {Solving the electronic {{Schr\"odinger}} equation
  for multiple nuclear geometries with weight-sharing deep neural networks},\
  }\href@noop {} {\bibfield  {journal} {\bibinfo  {journal} {arXiv:2105.08351
  [physics]}\ } (\bibinfo {year} {2021})},\ \Eprint
  {https://arxiv.org/abs/2105.08351} {arXiv:2105.08351 [physics]} \BibitemShut
  {NoStop}%
\bibitem [{\citenamefont {Neuscamman}\ \emph {et~al.}(2012)\citenamefont
  {Neuscamman}, \citenamefont {Umrigar},\ and\ \citenamefont
  {Chan}}]{neuscammanOptimizingLargeParameter2012}%
  \BibitemOpen
  \bibfield  {author} {\bibinfo {author} {\bibfnamefont {E.}~\bibnamefont
  {Neuscamman}}, \bibinfo {author} {\bibfnamefont {C.~J.}\ \bibnamefont
  {Umrigar}},\ and\ \bibinfo {author} {\bibfnamefont {G.~K.-L.}\ \bibnamefont
  {Chan}},\ }\bibfield  {title} {\bibinfo {title} {Optimizing large parameter
  sets in variational quantum {{Monte Carlo}}},\ }\href
  {https://doi.org/10.1103/PhysRevB.85.045103} {\bibfield  {journal} {\bibinfo
  {journal} {Physical Review B}\ }\textbf {\bibinfo {volume} {85}},\ \bibinfo
  {pages} {045103} (\bibinfo {year} {2012})},\ \Eprint
  {https://arxiv.org/abs/1108.0900} {arXiv:1108.0900} \BibitemShut {NoStop}%
\bibitem [{\citenamefont {Kingma}\ and\ \citenamefont
  {Ba}(2014)}]{kingmaAdamMethodStochastic2014}%
  \BibitemOpen
  \bibfield  {author} {\bibinfo {author} {\bibfnamefont {D.~P.}\ \bibnamefont
  {Kingma}}\ and\ \bibinfo {author} {\bibfnamefont {J.}~\bibnamefont {Ba}},\
  }\bibfield  {title} {\bibinfo {title} {Adam: A {{Method}} for {{Stochastic
  Optimization}}},\ }in\ \href@noop {} {\emph {\bibinfo {booktitle} {3rd
  {{International Conference}} for {{Learning Representations}}}}}\ (\bibinfo
  {year} {2014})\ \Eprint {https://arxiv.org/abs/1412.6980} {arXiv:1412.6980}
  \BibitemShut {NoStop}%
\bibitem [{\citenamefont {Huotari}\ \emph {et~al.}(2010)\citenamefont
  {Huotari}, \citenamefont {Soininen}, \citenamefont {Pylkk\"anen},
  \citenamefont {H\"am\"al\"ainen}, \citenamefont {Issolah}, \citenamefont
  {Titov}, \citenamefont {McMinis}, \citenamefont {Kim}, \citenamefont {Esler},
  \citenamefont {Ceperley}, \citenamefont {Holzmann},\ and\ \citenamefont
  {Olevano}}]{Na}%
  \BibitemOpen
  \bibfield  {author} {\bibinfo {author} {\bibfnamefont {S.}~\bibnamefont
  {Huotari}}, \bibinfo {author} {\bibfnamefont {J.~A.}\ \bibnamefont
  {Soininen}}, \bibinfo {author} {\bibfnamefont {T.}~\bibnamefont
  {Pylkk\"anen}}, \bibinfo {author} {\bibfnamefont {K.}~\bibnamefont
  {H\"am\"al\"ainen}}, \bibinfo {author} {\bibfnamefont {A.}~\bibnamefont
  {Issolah}}, \bibinfo {author} {\bibfnamefont {A.}~\bibnamefont {Titov}},
  \bibinfo {author} {\bibfnamefont {J.}~\bibnamefont {McMinis}}, \bibinfo
  {author} {\bibfnamefont {J.}~\bibnamefont {Kim}}, \bibinfo {author}
  {\bibfnamefont {K.}~\bibnamefont {Esler}}, \bibinfo {author} {\bibfnamefont
  {D.~M.}\ \bibnamefont {Ceperley}}, \bibinfo {author} {\bibfnamefont
  {M.}~\bibnamefont {Holzmann}},\ and\ \bibinfo {author} {\bibfnamefont
  {V.}~\bibnamefont {Olevano}},\ }\bibfield  {title} {\bibinfo {title}
  {Momentum distribution and renormalization factor in sodium and the electron
  gas},\ }\href {https://doi.org/10.1103/PhysRevLett.105.086403} {\bibfield
  {journal} {\bibinfo  {journal} {Phys. Rev. Lett.}\ }\textbf {\bibinfo
  {volume} {105}},\ \bibinfo {pages} {086403} (\bibinfo {year}
  {2010})}\BibitemShut {NoStop}%
\bibitem [{\citenamefont {Hiraoka}\ \emph {et~al.}(2020)\citenamefont
  {Hiraoka}, \citenamefont {Yang}, \citenamefont {Hagiya}, \citenamefont
  {Niozu}, \citenamefont {Matsuda}, \citenamefont {Huotari}, \citenamefont
  {Holzmann},\ and\ \citenamefont {Ceperley}}]{Li}%
  \BibitemOpen
  \bibfield  {author} {\bibinfo {author} {\bibfnamefont {N.}~\bibnamefont
  {Hiraoka}}, \bibinfo {author} {\bibfnamefont {Y.}~\bibnamefont {Yang}},
  \bibinfo {author} {\bibfnamefont {T.}~\bibnamefont {Hagiya}}, \bibinfo
  {author} {\bibfnamefont {A.}~\bibnamefont {Niozu}}, \bibinfo {author}
  {\bibfnamefont {K.}~\bibnamefont {Matsuda}}, \bibinfo {author} {\bibfnamefont
  {S.}~\bibnamefont {Huotari}}, \bibinfo {author} {\bibfnamefont
  {M.}~\bibnamefont {Holzmann}},\ and\ \bibinfo {author} {\bibfnamefont
  {D.~M.}\ \bibnamefont {Ceperley}},\ }\bibfield  {title} {\bibinfo {title}
  {Direct observation of the momentum distribution and renormalization factor
  in lithium},\ }\href {https://doi.org/10.1103/PhysRevB.101.165124} {\bibfield
   {journal} {\bibinfo  {journal} {Phys. Rev. B}\ }\textbf {\bibinfo {volume}
  {101}},\ \bibinfo {pages} {165124} (\bibinfo {year} {2020})}\BibitemShut
  {NoStop}%
\bibitem [{\citenamefont {Olevano}\ \emph {et~al.}(2012)\citenamefont
  {Olevano}, \citenamefont {Titov}, \citenamefont {Ladisa}, \citenamefont
  {H\"am\"al\"ainen}, \citenamefont {Huotari},\ and\ \citenamefont
  {Holzmann}}]{ComptGw}%
  \BibitemOpen
  \bibfield  {author} {\bibinfo {author} {\bibfnamefont {V.}~\bibnamefont
  {Olevano}}, \bibinfo {author} {\bibfnamefont {A.}~\bibnamefont {Titov}},
  \bibinfo {author} {\bibfnamefont {M.}~\bibnamefont {Ladisa}}, \bibinfo
  {author} {\bibfnamefont {K.}~\bibnamefont {H\"am\"al\"ainen}}, \bibinfo
  {author} {\bibfnamefont {S.}~\bibnamefont {Huotari}},\ and\ \bibinfo {author}
  {\bibfnamefont {M.}~\bibnamefont {Holzmann}},\ }\bibfield  {title} {\bibinfo
  {title} {Momentum distribution and compton profile by the ab initio gw
  approximation},\ }\href {https://doi.org/10.1103/PhysRevB.86.195123}
  {\bibfield  {journal} {\bibinfo  {journal} {Phys. Rev. B}\ }\textbf {\bibinfo
  {volume} {86}},\ \bibinfo {pages} {195123} (\bibinfo {year}
  {2012})}\BibitemShut {NoStop}%
\bibitem [{\citenamefont {Yang}\ \emph {et~al.}(2020)\citenamefont {Yang},
  \citenamefont {Hiraoka}, \citenamefont {Matsuda}, \citenamefont {Holzmann},\
  and\ \citenamefont {Ceperley}}]{Licore}%
  \BibitemOpen
  \bibfield  {author} {\bibinfo {author} {\bibfnamefont {Y.}~\bibnamefont
  {Yang}}, \bibinfo {author} {\bibfnamefont {N.}~\bibnamefont {Hiraoka}},
  \bibinfo {author} {\bibfnamefont {K.}~\bibnamefont {Matsuda}}, \bibinfo
  {author} {\bibfnamefont {M.}~\bibnamefont {Holzmann}},\ and\ \bibinfo
  {author} {\bibfnamefont {D.~M.}\ \bibnamefont {Ceperley}},\ }\bibfield
  {title} {\bibinfo {title} {Quantum monte carlo compton profiles of solid and
  liquid lithium},\ }\href {https://doi.org/10.1103/PhysRevB.101.165125}
  {\bibfield  {journal} {\bibinfo  {journal} {Phys. Rev. B}\ }\textbf {\bibinfo
  {volume} {101}},\ \bibinfo {pages} {165125} (\bibinfo {year}
  {2020})}\BibitemShut {NoStop}%
\bibitem [{\citenamefont {McMillan}(1965)}]{McMillan}%
  \BibitemOpen
  \bibfield  {author} {\bibinfo {author} {\bibfnamefont {W.~L.}\ \bibnamefont
  {McMillan}},\ }\bibfield  {title} {\bibinfo {title} {Ground state of liquid
  ${\mathrm{he}}^{4}$},\ }\href {https://doi.org/10.1103/PhysRev.138.A442}
  {\bibfield  {journal} {\bibinfo  {journal} {Phys. Rev.}\ }\textbf {\bibinfo
  {volume} {138}},\ \bibinfo {pages} {A442} (\bibinfo {year}
  {1965})}\BibitemShut {NoStop}%
\bibitem [{\citenamefont {Liu}\ \emph {et~al.}(1974)\citenamefont {Liu},
  \citenamefont {Kalos},\ and\ \citenamefont {Chester}}]{LiuKalosChester1974}%
  \BibitemOpen
  \bibfield  {author} {\bibinfo {author} {\bibfnamefont {K.~S.}\ \bibnamefont
  {Liu}}, \bibinfo {author} {\bibfnamefont {M.~H.}\ \bibnamefont {Kalos}},\
  and\ \bibinfo {author} {\bibfnamefont {G.~V.}\ \bibnamefont {Chester}},\
  }\bibfield  {title} {\bibinfo {title} {Quantum hard spheres in a channel},\
  }\href {https://doi.org/10.1103/PhysRevA.10.303} {\bibfield  {journal}
  {\bibinfo  {journal} {Physical Review A}\ }\textbf {\bibinfo {volume} {10}},\
  \bibinfo {pages} {303} (\bibinfo {year} {1974})}\BibitemShut {NoStop}%
\bibitem [{\citenamefont {Baroni}\ and\ \citenamefont {Moroni}(1999)}]{rept}%
  \BibitemOpen
  \bibfield  {author} {\bibinfo {author} {\bibfnamefont {S.}~\bibnamefont
  {Baroni}}\ and\ \bibinfo {author} {\bibfnamefont {S.}~\bibnamefont
  {Moroni}},\ }\bibfield  {title} {\bibinfo {title} {Reptation quantum monte
  carlo: A method for unbiased ground-state averages and imaginary-time
  correlations},\ }\href {https://doi.org/10.1103/PhysRevLett.82.4745}
  {\bibfield  {journal} {\bibinfo  {journal} {Phys. Rev. Lett.}\ }\textbf
  {\bibinfo {volume} {82}},\ \bibinfo {pages} {4745} (\bibinfo {year}
  {1999})}\BibitemShut {NoStop}%
\bibitem [{\citenamefont {Ceperley}\ and\ \citenamefont
  {Kalos}(1979)}]{extrapol}%
  \BibitemOpen
  \bibfield  {author} {\bibinfo {author} {\bibfnamefont {D.}~\bibnamefont
  {Ceperley}}\ and\ \bibinfo {author} {\bibfnamefont {M.}~\bibnamefont
  {Kalos}},\ }\href@noop {} {\emph {\bibinfo {title} {Quantum Many-Body
  Problems in Monte Carlo Methods in Statistical Physics}}},\ edited by\
  \bibinfo {editor} {\bibfnamefont {K.}~\bibnamefont {Binder}}\ (\bibinfo
  {publisher} {Springer-Verlag},\ \bibinfo {year} {1979})\BibitemShut {NoStop}%
\bibitem [{\citenamefont {Holzmann}\ \emph {et~al.}(2016)\citenamefont
  {Holzmann}, \citenamefont {Clay~III}, \citenamefont {Morales}, \citenamefont
  {Tubman}, \citenamefont {Ceperley},\ and\ \citenamefont {Pierleoni}}]{fse}%
  \BibitemOpen
  \bibfield  {author} {\bibinfo {author} {\bibfnamefont {M.}~\bibnamefont
  {Holzmann}}, \bibinfo {author} {\bibfnamefont {R.~C.}\ \bibnamefont
  {Clay~III}}, \bibinfo {author} {\bibfnamefont {M.~A.}\ \bibnamefont
  {Morales}}, \bibinfo {author} {\bibfnamefont {N.~M.}\ \bibnamefont {Tubman}},
  \bibinfo {author} {\bibfnamefont {D.~M.}\ \bibnamefont {Ceperley}},\ and\
  \bibinfo {author} {\bibfnamefont {C.}~\bibnamefont {Pierleoni}},\ }\bibfield
  {title} {\bibinfo {title} {Theory of finite size effects for electronic
  quantum monte carlo calculations of liquids and solids},\ }\href@noop {}
  {\bibfield  {journal} {\bibinfo  {journal} {Physical Review B}\ }\textbf
  {\bibinfo {volume} {94}},\ \bibinfo {pages} {035126} (\bibinfo {year}
  {2016})}\BibitemShut {NoStop}%
\bibitem [{Rea(1967)}]{ReattoChester1967}%
  \BibitemOpen
  \bibfield  {title} {\bibinfo {title} {Phonons and the properties of a bose
  system},\ }\href {https://doi.org/10.1103/PhysRev.155.88} {\bibfield
  {journal} {\bibinfo  {journal} {Phys. Rev.}\ }\textbf {\bibinfo {volume}
  {155}},\ \bibinfo {pages} {88} (\bibinfo {year} {1967})}\BibitemShut
  {NoStop}%
\bibitem [{\citenamefont {Moroni}\ \emph {et~al.}(1997)\citenamefont {Moroni},
  \citenamefont {Senatore},\ and\ \citenamefont {Fantoni}}]{SaverioMomkHe}%
  \BibitemOpen
  \bibfield  {author} {\bibinfo {author} {\bibfnamefont {S.}~\bibnamefont
  {Moroni}}, \bibinfo {author} {\bibfnamefont {G.}~\bibnamefont {Senatore}},\
  and\ \bibinfo {author} {\bibfnamefont {S.}~\bibnamefont {Fantoni}},\
  }\bibfield  {title} {\bibinfo {title} {Momentum distribution of liquid
  helium},\ }\href {https://doi.org/10.1103/PhysRevB.55.1040} {\bibfield
  {journal} {\bibinfo  {journal} {Phys. Rev. B}\ }\textbf {\bibinfo {volume}
  {55}},\ \bibinfo {pages} {1040} (\bibinfo {year} {1997})}\BibitemShut
  {NoStop}%
\bibitem [{\citenamefont {Lin}\ \emph {et~al.}(2001)\citenamefont {Lin},
  \citenamefont {Zong},\ and\ \citenamefont {Ceperley}}]{lin2001twist}%
  \BibitemOpen
  \bibfield  {author} {\bibinfo {author} {\bibfnamefont {C.}~\bibnamefont
  {Lin}}, \bibinfo {author} {\bibfnamefont {F.}~\bibnamefont {Zong}},\ and\
  \bibinfo {author} {\bibfnamefont {D.~M.}\ \bibnamefont {Ceperley}},\
  }\bibfield  {title} {\bibinfo {title} {Twist-averaged boundary conditions in
  continuum quantum monte carlo algorithms},\ }\href@noop {} {\bibfield
  {journal} {\bibinfo  {journal} {Physical Review E}\ }\textbf {\bibinfo
  {volume} {64}},\ \bibinfo {pages} {016702} (\bibinfo {year}
  {2001})}\BibitemShut {NoStop}%
\bibitem [{\citenamefont {Cassella}\ \emph {et~al.}(2023)\citenamefont
  {Cassella}, \citenamefont {Sutterud}, \citenamefont {Azadi}, \citenamefont
  {Drummond}, \citenamefont {Pfau}, \citenamefont {Spencer},\ and\
  \citenamefont {Foulkes}}]{PhysRevLett.130.036401}%
  \BibitemOpen
  \bibfield  {author} {\bibinfo {author} {\bibfnamefont {G.}~\bibnamefont
  {Cassella}}, \bibinfo {author} {\bibfnamefont {H.}~\bibnamefont {Sutterud}},
  \bibinfo {author} {\bibfnamefont {S.}~\bibnamefont {Azadi}}, \bibinfo
  {author} {\bibfnamefont {N.~D.}\ \bibnamefont {Drummond}}, \bibinfo {author}
  {\bibfnamefont {D.}~\bibnamefont {Pfau}}, \bibinfo {author} {\bibfnamefont
  {J.~S.}\ \bibnamefont {Spencer}},\ and\ \bibinfo {author} {\bibfnamefont
  {W.~M.~C.}\ \bibnamefont {Foulkes}},\ }\bibfield  {title} {\bibinfo {title}
  {Discovering quantum phase transitions with fermionic neural networks},\
  }\href {https://doi.org/10.1103/PhysRevLett.130.036401} {\bibfield  {journal}
  {\bibinfo  {journal} {Phys. Rev. Lett.}\ }\textbf {\bibinfo {volume} {130}},\
  \bibinfo {pages} {036401} (\bibinfo {year} {2023})}\BibitemShut {NoStop}%
\bibitem [{\citenamefont {Li}\ \emph {et~al.}(2022{\natexlab{b}})\citenamefont
  {Li}, \citenamefont {Li},\ and\ \citenamefont {Chen}}]{LiLi}%
  \BibitemOpen
  \bibfield  {author} {\bibinfo {author} {\bibfnamefont {X.}~\bibnamefont
  {Li}}, \bibinfo {author} {\bibfnamefont {Z.}~\bibnamefont {Li}},\ and\
  \bibinfo {author} {\bibfnamefont {J.}~\bibnamefont {Chen}},\ }\bibfield
  {title} {\bibinfo {title} {Ab initio calculation of real solids via neural
  network ansatz},\ }\href {https://doi.org/10.1038/s41467-022-35627-1}
  {\bibfield  {journal} {\bibinfo  {journal} {Nature Communications}\ }\textbf
  {\bibinfo {volume} {13}},\ \bibinfo {pages} {7895} (\bibinfo {year}
  {2022}{\natexlab{b}})}\BibitemShut {NoStop}%
\end{thebibliography}%
